\documentclass[journal]{IEEEtran}
\IEEEoverridecommandlockouts
\usepackage{cite}
\usepackage{amsmath,amssymb,amsfonts,amsthm}
\usepackage{algorithmic}
\usepackage{graphicx}
\usepackage{textcomp}
\usepackage{xcolor}
\usepackage{float}
\usepackage{dotlessi}
\usepackage{tcolorbox}
\usepackage{subcaption}
\usepackage{mathtools}
\usepackage{caption}
\usepackage{cuted}
\def\BibTeX{{\rm B\kern-.05em{\sc i\kern-.025em b}\kern-.08em
    T\kern-.1667em\lower.7ex\hbox{E}\kern-.125emX}}

\newcommand{\M}[4]{
\renewcommand{\arraystretch}{0.8}
\setlength{\arraycolsep}{2pt}
\fontsize{7}{12}\selectfont
\begin{bmatrix}#1 & #2\\ #3 & #4\end{bmatrix}}
\newcommand{\bfg}[1]{\boldsymbol{#1}}
\newcommand{\ul}[1]{\underline{#1}}
\newcommand{\ull}[1]{\underline{\underline{#1}}}
\newcommand{\vvec}{\bar{\bfg v}}

\newcommand{\ivec}{\bar{\bfg \imath}}
\newcommand{\Ivec}{\bar{\bfg \imath}_{{\rm dev}}}
\newcommand{\dIvec}{\dot{\bar{\bfg \imath}}_{{\rm dev}}}
\newcommand{\ddIvec}{\ddot{\bar{\bfg \imath}}_{{\rm dev}}}

\newcommand{\idev}{\bar{\imath}_{\rm dev}}
\newcommand{\idevdot}{\dot{\bar{\imath}}_{\rm dev}}
\newcommand{\idevddot}{\ddot{\bar{\imath}}_{\rm dev}}
\newcommand{\videv}{\ul{\imath}_{\rm dev}}
\newcommand{\videvdot}{\dot{\ul{\imath}}_{\rm dev}}
\newcommand{\videvddot}{\ddot{\ul{\imath}}_{\rm dev}}

\newcommand{\vivec}{\ul{\bfg \imath}}
\newcommand{\vIvec}{\ul{\bfg \imath}_{\bfg{\rm dev}}}
\newcommand{\vdIvec}{\dot{\ul{\bfg \imath}}_{\bfg{\rm dev}}}
\newcommand{\vddIvec}{\ddot{\ul{\bfg \imath}}_{\bfg{\rm dev}}}

\newcommand{\vdv}{\Delta\ul{\bfg v}}
\newcommand{\vdvdot}{\Delta\dot{\ul{\bfg v}}}
\newcommand{\vdvddot}{\Delta\ddot{\ul{\bfg v}}}

\newcommand{\Iden}{\ull{\boldsymbol{\rm I}}}

\newcommand{\Zmat}{\bar{\bfg Z}}
\newcommand{\mZmat}{\ull{\bfg Z}}
\newcommand{\mZmateq}{\ull{\bfg Z_{\rm eq}}}
\newcommand{\mZmateqp}{\ull{\bfg Z_{\rm eq}'}}
\newcommand{\mZmateqpp}{\ull{\bfg Z_{\rm eq}''}}

\newcommand{\uletavecam}{\widetilde{\ull{\bfg \eta}}'}
\newcommand{\uletaavecam}{\widetilde{\ull{\bfg \eta}}''}

\newcommand{\etilde}{\widetilde{\ull{{\rm e}}^{\, \jmath\,\theta}}}
\newcommand{\etheta}{{\rm e}^{\, \jmath\,\theta}}

\newcommand{\dv}{\Delta \bar{\bfg v}}
\newcommand{\di}{\Delta \bar{\bfg \imath}}

\newcommand{\szero}{\ull{\bfg S}}
\newcommand{\sone}{\ull{\bfg S}'}
\newcommand{\stwo}{\ull{\bfg S}''}
\newcommand{\dvs}{\Delta\ul{\bfg v}_{\rm v\theta}}
\newcommand{\detas}{\Delta\ul{\bfg \eta}'}
\newcommand{\detaas}{\Delta\ul{\bfg \eta}''}
\newcommand{\dis}{\Delta\ul{\bfg \imath}_{\rm pq}}

\begin{document}

\title{Analytical Framework for Power System Strength} % - Part I: Theory}

\author{Ignacio Ponce,~\IEEEmembership{Graduate Student Member,~IEEE,}
        and Federico Milano,~\IEEEmembership{Fellow,~IEEE}%
\thanks{I. Ponce and F. Milano are with the School of Electrical and Electronic Engineering, University College Dublin, Belfield Campus, D04V1W8, Ireland.
e-mails: ignacio.poncearancibia@ucdconnect.ie, federico.milano@ucd.ie}%
\thanks{This work is supported by Sustainable Energy Authority of Ireland (SEAI) by funding I.  Ponce and F.  Milano under project FRESLIPS, Grant No.~RDD/00681.}%
\vspace{-7mm}
}

\maketitle

\begin{abstract}
%This two-part series article introduces a general formulation for evaluating power system strength.  This paper is the first of the series and focuses on its theoretical foundations.  
This paper proposes a general framework to evaluate power system strength.  
% Strength is conceived as a property of the bus voltage phasors and their derivatives that reflect their ``resistance'' to change following perturbations.  
The formulation features twelve indicators, grouped in three dynamical orders, that quantify the resistance of bus voltage phasors and their first and second order rates of change to sudden current injection changes.  To quantify such changes, the paper introduces a novel finite differentiation technique, which we named \textit{Delta operator}, able to properly capture ``jumps'' of algebraic variables and utilizes the recently developed concept of complex frequency.  
% A key mathematical byproduct, the Delta operator, is defined and characterized.  
% The formulation is analytical and substantially more general than existing strength metrics, unifying traditional voltage and frequency strength indicators within a single framework.  
The paper also shows how the proposed framework can be systematically applied to any system device, and provides a variety of examples based on synchronous machines, converters, and loads models are given.  Numerical results in a benchmark system validate the exactness of the formulation.  
% Limitations, applications and other practical aspects of the proposal are addressed in the second part of the series.
\end{abstract}

\vspace{-2mm}
\section{Introduction}

\subsection{Motivation}

% In power systems, `strength' is a concept that has gained increasing attention in the last decades due to its importance for ensuring an adequate operation of modern power systems.  Although it lacks a standard definition, there is a broad consensus on its essential meaning.  

The term `strength' refers to the resistance of a system to disturbances.  Intuitively, a `stronger' system is less sensitive to perturbations, whereas a `weaker' system experiences higher deviations when subjected to the same disturbance \cite{cigrestrength}.  
% In practice, strength quantifies the resistance of system output variables to perturbations at a point of the network.  
% Although many variables can be considered,
Typically, the voltage is the representative variable for evaluating strength.  %, where `voltage' refers to magnitude, angle and frequency.  
% the multidimensional electrical variable at the bus, not a single scalar such as the magnitude.  
For instance, the Australian Energy Market Commission (AEMC) defines system strength as \textit{the power system's ability to resist the changes in the magnitude, phase angle, and waveform of the voltage at any given location under different operating conditions} \cite{aemc}.
%
% Voltage strength can be quantified following different methodologies, which can be broadly classified into two categories: data-driven, or analytical methods.  Data-driven techniques use actual measurements, simulation results, and other data sources to estimate voltage strength.  Some recent examples can be found in \cite{forecastingstrength,wangqi,boricic,huang1,huang2}.  In turn, analytical approaches provide indicators for quantifying voltage strength as a function of variables and/or parameters of a power system model.
%
How to quantify this ability to resist changes, however, is currently an open question.
% To answer this question,
In this paper, we propose a general analytical framework to evaluate `strength' based on a novel finite-difference operator specifically developed for this purpose.

\vspace{-2mm}
\subsection{Literature review}

% In alternating current (AC) three-phase systems, 

Analytical developments for quantifying voltage strength are separated into specific indicators to assess voltage magnitude strength, and separate metrics to evaluate frequency strength.  This separation establishes a natural relationship with conventional types of stability analysis, namely rotor angle, frequency, and voltage stability \cite{taskforce1}.

The foundation of voltage magnitude strength is the short-circuit level (SCL) at a given location \cite{kundur}.  %, also referred as the fault level.  
% It is well-recognized the close relationship between the SCL and the sensitivity of the voltage magnitude to power flow changes \cite{kundur}.
%The SCL is commonly approximated as the inverse of the Thevenin impedance at a bus, whose reactance is normally much higher than the resistance in high voltage transmission networks, making the voltage magnitude much more sensitive to reactive power changes than to active power changes 
The SCL has historically been used as a measure of strength and has become a critical factor in ensuring the stable operation of modern systems with inverter-based resources (IBRs).  In particular, the short-circuit ratio (SCR), defined as the ratio of the SCL to the rated power of the IBR, has emerged as a basic metric to assess the ability of the IBR to withstand low strength conditions \cite{teodorescu}.  
% Typically, SCR values above three are considered strong, and weak values range between three and two or even below two \cite{wuscrval}.  
The SCR ignores the effect of neighboring converters and can lead to optimistic results in networks with multiple IBRs.  This issue has motivated the proposal of the weighted SCR \cite{wscr}, the composite SCR \cite{cscr}, the Generalized SCR (gSCR), and a wide list of more elaborated indicators addressing the shortcomings of the SCR \cite{iscr,sdscr,gsimp,task2018force,kim2020,gscr,gscr2}.  A comparison between these metrics can be found in \cite{scrreview}.

With regard to frequency, strength metrics are focused on the system dynamic performance.  %have evolved in a different manner in the case of the frequency.  
Conventional AC systems are dominated by synchronous machines (SMs), which operate by nature at a unique frequency, rendering it a rather global variable of the grid.  Consequently, the principles of frequency dynamics have been studied based on simplified single-node models or using global averaged quantities, such as the frequency of the center of inertia.  This scenario has also led to the use of a global metric for quantifying frequency strength, namely the \textit{system inertia}, defined as the sum of the inertia of the SMs.  
% Inertia is a core property of SMs that massively affects their dynamic behavior, as manifested in the so-called swing equation \cite{kundur}.  
It is a common assumption that low inertia levels combined with a high penetration of IBRs endanger power systems' operation.  This is a very common scenario in power systems nowadays, and has given rise to several challenges for frequency control, and created a paradigm shift in the study of frequency dynamics \cite{foundations}.  

Recent investigations have shown that the spatial distribution of inertia significantly impacts the post-disturbance local frequency response \cite{ulbig,osbouei,adrees}.  These observations have motivated research on novel indicators aimed at reflecting local frequency strength, quantifying the inertia concentration at different locations of the network 
\cite{brahma, warren, zeng, gosh, pinheiro, idenareas}.
Despite notable methodological differences, most works are based on the so-called `nodal inertia', whose foundation assumes the existence of an equivalent swing equation at buses \cite{zeng}.  While some of these proposal's ultimate goal is to calculate this nodal-level metric \cite{zeng, gosh,pinheiro}, others use it as an intermediate step to get regional frequency indicators, recognizing frequency coherency in areas of the grid \cite{idenareas}.

Current research efforts on strength metrics still rely on strong approximations of the system representation, and often involve equations proposed empirically rather than derived analytically.  In addition, indicators for quantifying the voltage magnitude strength and voltage frequency strength are derived independently with fundamentally different approaches, even though both variables are, ultimately, two components of a unique entity: a three-phase AC voltage.  Recent developments have enabled the search for a more general and unifying formulation.  In particular, the complex frequency (CF) is a concept recently proposed in \cite{ComplexFreq} that has given rise to a variety of improvements in power system modeling and control \cite{complexprop, complexacdc, cfd, bernal2025cf, poncelocalsynch}.  This article exploits the properties of the CF to develop the proposed analytical framework to evaluate system strength.
% for strength indicators.
% to facilitate the mathematical representation of the voltage and frequency and the relationship between these quantities.

%Even in the case of power systems that remain conventional nowadays (HV systems highly based on synchronous generation), a more general formulation over which assumptions can be taken and recover the particular cases appears a more adequate solution.

% While the magnitude strength calculation is based on short-circuit levels and topology-based metrics, the frequency strength is measured in terms of inertia, i.e., the time constant of a first-order differential equation.

\subsection{Contribution}

The contributions of the paper are threefold.
\begin{itemize}
\item A novel general analytical framework to evaluate power system strength in steady-state and dynamic conditions.  
% The framework substantially more general than the current state of art.  It allows evaluating the strength of the different components of the voltage---conceived as a multidimensional electrical variable at a given location--- in a consistent manner.  
\item A systematic methodology to study the effect of diverse device models on system strength.  Specific expressions for relevant devices, such as synchronous machines, converters, and loads, are provided.
\item % A notable byproduct resulting from the development requirements of the formulation is the 
Definition of a novel mathematical operator, called \textit{Delta operator}, along with some of its properties and identities.
\end{itemize}
A relevant consequence of the proposed analytical framework is the unification of the voltage magnitude and frequency strength assessment.  It also gives further information on the strength of buses when subjected to remote perturbations.  The proposed framework is purely analytical, minimizes assumptions on the system representation, and avoids approximations, rendering it `exact', provided the system model is exact.

\subsection{Paper organization}

% This paper is the first of a two-part series.  Part I provides the theoretical foundations of the proposed formulation of system strength, and Part II delves into a variety of applications and practical aspects of the proposal in real-world scenarios.

The remainder of this document is organized as follows.  Section \ref{sec:mathematical_background} introduces key mathematical foundations required for the analytical developments of the paper.  Section \ref{sec:generalization_of_system_strength} introduces the proposed formulation of system strength and its derivation.  Section \ref{sec:devmodels} presents specific expressions for basic power system devices that are relevant to system strength.  Section \ref{sec:study_case} offers study cases implementing the proposal in benchmark systems.  Finally, Section \ref{sec:conclusion} draws the main conclusions and outlines future work.  
% main contributions andthe transition to Part II and concludes the paper.

\section{Mathematical background}
\label{sec:mathematical_background}

This section presents the mathematical foundations required to derive the system strength indicators presented in this work.  

\subsection{Complex frequency}
\label{subsec:complex_frequency} 

Consider the voltage represented as a dynamic Clarke vector $\bar{v}\in \mathbb{C}\,|\,\bar{v}=v\cos\theta+\jmath\, v\sin\theta$.  The complex frequency (CF) of the voltage is a complex quantity denoted as $\bar{\eta}$ and defined in \cite{ComplexFreq} as follows:
\begin{equation}
    \bar{\eta}=\frac{\dot{v}}{v}+\jmath\, \dot{\theta}=\rho+\jmath\,\omega,\quad v\neq0\,.
\end{equation}

The CF is related to the total time derivative of the voltage as in the equation below:
\begin{equation}\label{eq:cf_prop1}
    \bar{p}\{\bar{v}\}=\bar{v}\,\bar{\eta}\,,
\end{equation}
where $\bar{p}\{\cdot\}=\frac{d}{dt}\{\cdot\}+\{\cdot\}\,\jmath\,\dot{\theta}_{\rm dq}$ is the total derivative operator of a dynamic Clarke vector as defined in \cite{milanofrequency}, and $\theta_{\rm dq}$ the reference angle of the $\rm dq$ coordinates with respect to a fixed reference angle.  Therefore, \eqref{eq:cf_prop1} gives the `absolute' time derivative of $\bar{v}$, i.e., the rate of change of the vector relative to the fixed reference frame, independently of the $\rm dq$ coordinates used to represent $\bar{v}$.  We are interested in calculating the time derivative of the vector relative to a rotating reference frame.  To do so, \eqref{eq:cf_prop1} becomes:
\begin{equation}\label{eq:cf_prop1_5}
    \dot{\bar{v}}=\bar{v}\,\left(\bar{\eta}-\jmath\,\omega_{0}\right)\,,
\end{equation}
where $\omega_0$ is the fundamental frequency of the system.

Equation \eqref{eq:cf_prop1_5} is a very useful property of the CF that allows using it as a time derivative operator.  In this work, we are also interested in calculating the second-order time derivative of the voltage, $\ddot{\bar{v}}$, for which a quantity with a property similar to \eqref{eq:cf_prop1_5} would highly facilitate the formulation.  This motivates defining a second-order CF, $\bar{\eta}''$, as the complex quantity satisfying the following equation:
\begin{equation}\label{eq:cf_prop2}
    \ddot{\bar{v}}=\bar{v}\,\bar{\eta}'',
\end{equation}
An expression for $\bar{\eta}''$ is found by taking \eqref{eq:cf_prop1_5} and applying the time derivative at both sides:
\begin{align}
    \ddot{\bar{v}}&=\dot{\bar{v}}\,(\bar{\eta}-\jmath\,\omega_0)+\bar{v}\,\dot{\bar{\eta}}\\
    \Leftrightarrow\ddot{\bar{v}}&=\bar{v}\,\left((\bar{\eta}-\jmath\,\omega_0)^2+\dot{\bar{\eta}}\right)\\
\Rightarrow \bar{\eta}'' &= (\bar{\eta}-\jmath\,\omega_0)^2+\dot{\bar{\eta}}\,.
\end{align}

The real and imaginary parts of $\bar{\eta}''$ are hereafter denoted as $\sigma$ and $\gamma$, respectively, i.e., $\bar{\eta}''=\sigma+\jmath\,\gamma$, where:
\begin{align}
    \sigma &= \rho^2-(\omega-\omega_0)^2+\dot{\rho}\,;\quad \gamma = 2\rho\,(\omega-\omega_0)+\dot{\omega}\,.\label{eq:gammadef}
\end{align} 

In the remainder of this work, the (original) first-order relative CF (i.e., $\bar{\eta}-\jmath\,\omega_0$) is denoted as $\bar{\eta}'$ and the second-order CF is denoted as $\bar{\eta}''$.  In practice, $\sigma\approx \dot{\rho}$ and $\gamma\approx \dot{\omega}$, implying that the second-order CF closely represents the rate of change of the first-order CF. Nevertheless, for this paper, it is relevant to recognize that $\bar{\eta}''$ contains the information of the second-order dynamics of the voltage vector, similarly to how the CF packs the first-order dynamics of the vector.  Providing a complete physical interpretation of this quantity is out of the scope of this work.

\subsection{Delta operator}

Consider a standard dynamic model of power systems in the form of a set of differential algebraic equations (DAEs).  Let $f(t)$ be a scalar function of time $f(t):\mathbb{R}^{+}\rightarrow\mathbb{R}$ representing an arbitrary variable of the set of DAEs, which can be an input, state, or algebraic variable.  Note $f$ might have discontinuities in the latter case as algebraic variables can \textit{jump} at specific times, e.g., due to the ocurrence of faults.

\textit{Definition 1:} Delta ($\Delta$) operator applied to $f(t)$:
\begin{equation}\label{eq:def_delta}
  \Delta f(t):=\lim_{\tau\to t^{+}}f(\tau)-\lim_{\tau\to t^{-}}f(\tau)\,.  
\end{equation}

In simple words, $\Delta f(t)$ gives the difference between the value of $f$ evaluated at a time infinitesimally after $t$ ($t^{+}$) and infinitesimally before $t$ ($t^{-}$).  Note $\Delta f(t)$ is always null unless $f$ is discontinuous at $t$.  Hereafter, the notation used for the limits of the function when $\tau$ approaches $t^{+}$ and $t^{-}$ is simplified as $f^{+}$ and $f^{-}$, respectively.  Thus:
\begin{equation}\label{eq:def_delta_simp}
    \Delta f(t)= f^{+} - f^{-}\,.
\end{equation}

\textit{Definition 2:} The \textit{instantaneous arithmetic mean} of $f$:
\begin{equation}\label{eq:def_iam}
    \widetilde{f}(t):=\frac{f^{+} + f^{-}}{2}\,.
\end{equation}

\textit{Definition 3:} The \textit{instantaneous geometric mean} of $f$:
\begin{equation}\label{eq:def_igm}
    \hat{f}(t):=\sqrt{f^{+} f^{-}}\,.
\end{equation}

Let $f(t)$, $g(t)$ be variables of the set of DAEs, and $\alpha,\beta$ constants.  The $\Delta$ operator satisfies the properties presented below, whose proofs can be found in the addendum provided with the paper.

\textit{Property 1}: $\Delta$ of a constant with time is null.
\begin{equation}
    \Delta\alpha=0\,.
\end{equation}

\textit{Property 2}: Linearity.
\begin{equation}
    \Delta\{\alpha f(t)+\beta g(t)\}=\alpha\Delta f(t)+\beta \Delta g(t)\,.
\end{equation}

\textit{Property 3}: Multiplication rule.
\begin{equation}
    \Delta\{f(t)g(t)\}=\Delta f(t)\widetilde{g}(t)+\widetilde{f}(t)\Delta g(t)\,.
\end{equation}

\textit{Property 4}: Division rule.
\begin{equation}
    \Delta\left\{\frac{f(t)}{g(t)}\right\}=\frac{\Delta f(t)\widetilde{g}(t)-\widetilde{f}(t)\Delta g(t)}{\hat{g}(t)^2}\,.
\end{equation}

\textit{Property 5:} Chain rule of the complex exponential function:
%\begin{equation}\label{eq:prop5}
%    \Delta e^{\jmath\, f(t)}=e^{\jmath\, \widetilde{f}(t)}\frac{\sin(\Delta f(t)/2)}{\Delta f(t)/2}\jmath\, \Delta f(t)\,.
%\end{equation}
%
%An alternative form of \eqref{eq:prop5} is shown below:
\begin{equation}\label{eq:prop5_alt}
    \Delta e^{\jmath\, f(t)}=\widetilde{e^{\jmath\, f(t)}}\jmath\,\frac{\tan(\Delta f(t)/2)}{1/2}\,.
\end{equation}

In case we calculate the limit when $f^{+}\to f^{-}$ of the definitions and properties stated above, the $\Delta$ operator becomes equivalent to the absolute derivative operator $(d)$, i.e., $\lim_{f^{+}\to f^{-}}\Delta f(t)=df(t)$.  The proof for properties 1 and 2 is trivial.  It also comes straightforwardly for properties 3 and 4 since $\lim_{f^{+}\to f^{-}}\widetilde{f}(t)=\lim_{f^{+}\to f^{-}}\hat{f}(t)=f(t)$. Finally, the proof of property 5 needs recalling that $\lim_{x\to0}\,\sin(x)/x=1$. This highlights the consistency of the definition and properties found for $\Delta$.

Based on the properties above, some identities regarding the use of the $\Delta$ operator with Clarke vectors are found.  Hereafter, the time dependency $(t)$ is omitted for simplicity.

\textit{Identity 1:} $\Delta$ of a Clarke vector $\bar{v}=v\,\etheta$:
%\begin{equation}\label{eq:identity1}
%    \Delta\bar{v}=\widetilde{v}e^{\jmath\,\widetilde{\theta}}\left(\frac{\Delta v}{\widetilde{v}}\cos(\Delta \theta /2 )+\jmath\,\Delta\theta\,\frac{\sin(\Delta\theta/2)}{\Delta\theta/2}\right)
%\end{equation}
%An alternative form of \eqref{eq:identity1} is shown below:
\begin{equation}
\label{eq:identity1_alt}
\Delta\bar{v}=\widetilde{v}\widetilde{{\rm e}^{\, \jmath\,\theta}}\left(\frac{\Delta v}{\widetilde{v}}+\jmath\,\frac{\tan(\Delta\theta/2)}{1/2}\right) .
\end{equation}

\textit{Identity 2:} Consider a Clarke vector $\bar{\imath}$ and its equivalent after applying the Park transform with an angle $\theta_{\rm dq}$, $\bar{\imath}_{\rm dq}=\bar{\imath}e^{-\jmath\,\theta_{\rm dq}}$.  The following property holds after applying $\Delta$ to the original vector:
\begin{equation}\label{eq:identity2}
    \Delta\bar{\imath}=\widetilde{e^{\jmath\,\theta_{\rm dq}}}\left(\Delta \bar{\imath}_{\rm dq}+\widetilde{\bar{\imath}}_{\rm dq}\,\jmath\,\frac{\tan(\Delta\theta_{\rm dq}/2)}{1/2}\right) .
\end{equation}

The definitions, properties, and identities given in this section are, to the best of our knowledge, a novelty of this paper.

\section{Generalization of system strength}\label{sec:generalization_of_system_strength}
\subsection{Preliminaries}\label{subsec:strength_intro}
We aim to provide a more general set of indicators to quantify system strength.  As mentioned in the introduction, given the broad nature of the concept, we start by introducing the following considerations to treat system strength:
\begin{itemize}
    \item It is conceived as a property of each bus of the network.
    \item The voltage at the bus is taken as the representative variable over which strength is to be quantified.
    \item The voltage strength is evaluated with respect to changes in the current injected by a fictitious independent current source at the bus.
    \item The concept of strength is not restricted to a sensitivity in a small-signal sense, namely, variations of the voltage due to infinitesimal current changes (in which case we would look at a derivative-like expression such as $\frac{dV}{d\imath}$).  In turn, strength is a measure applicable to large-signal or discrete events (e.g., $\frac{\Delta V}{\Delta \imath}$).
\end{itemize}

\subsection{Proposed formulation}
\label{subsec:strength_proposal}

% This section presents the desired form of the equations needed for a general formulation to assess voltage strength.  

The usual approach to define system strength is to establish a relationship between the sensitivity of the voltage vector to a sudden change in the current injection at the bus.  Such a formulation can only take into account how the voltage vector changes instantaneously due to the current jump.  However, recalling the essence of the concept of strength as the voltage's resistance to perturbations, such an approach falls short in representing strength. It cannot evaluate how much the voltage will evolve in time right after the disturbance. We consider this an important aspect of the evaluation of strength.

Thus, we aim at a more general formulation to evaluate voltage strength that is able to reveal: (i) how much the voltage is expected to jump, and (ii) how fast it will continue to deviate right after the current change.  
% In practice, the latter information is contained in the voltage time derivatives.  In this work, 
Specifically, we propose a formulation composed of three categories of indicators, depending on the order of the time derivative involved: zero, first, and second-order strength.

\subsubsection{Zero-order strength}

Sensitivity of the voltage vector $\Delta\bar{V}$ to changes in the current of the independent source $\Delta\bar{\imath}$.  As $\bar{V}$ and $\bar{\imath}$ are 2-dimensional quantities, the sensitivity is 4-dimensional, i.e., there are four sensitivity factors corresponding to the possible combinations between the two components used to represent each vector.  These components must be chosen and may differ for the voltage and current.  Without lack of generality, the coordinates chosen for the independent current source are its active and reactive components ($\imath_{\rm p}$, $\imath_{\rm q}$), namely, the component that is in phase and in quadrature with the voltage vector, respectively.  In turn, we use polar coordinates for the voltage ($v$, $\theta$).  Therefore, ideally, we would evaluate the sensitivity of $v$ and $\theta$ to changes in $\imath_{\rm p}$ and $\imath_{\rm q}$.  However, note the voltage components have different dimensions: $volt$ and $rad/s$, which would make the strength indicators of different dimensions as well, complicating their interpretation and use.  For this reason, we formulate the strength indicators over functions of $v$ and $\theta$ that avoid this problem.  We formalize this formulation using a matrix representation as shown below:
\begin{align}
        \left[\begin{array}{cc}\Delta v/\widetilde{v}\\2\tan(\Delta\theta/2)\end{array}\right]&=
        \left[\begin{array}{cc}S_{v\imath_{\rm p}}&S_{v\imath_{\rm q}}\label{eq:sdef00} \\
        S_{\theta\imath_{\rm p}}&S_{\theta\imath_{\rm q}}\end{array}\right]
        \left[\begin{array}{cc}\Delta \imath_{\rm p}\\\Delta \imath_{\rm q}\end{array}\right] , \\
        \Delta\underline{v}_{\rm v\theta}&=
        \ull{S} \, \Delta \underline{\imath}_{\rm pq} \, . \label{eq:predictor0}
\end{align}

The magnitude of the voltage is normalized by its instantaneous arithmetic mean.  In the case of the angle, note the tangent is a bijective function for $\Delta\theta\in\left(-\pi,\pi\right)$, and for small angles in a vicinity of zero $2\tan(\Delta\theta/2)\approx\Delta\theta$.  Hereafter, subscript $\rm v\theta$ denotes the voltage vector parametrized using these functions.  The convenience of the form chosen will be evident in the derivation presented later in Section \ref{subsec:derivation}.

\subsubsection{First-order strength}
Sensitivity of the first-order dynamic of the voltage vector, represented through the first-order CF $\Delta\bar{\eta}'$, in real and imaginary parts ($\rho$, $\omega$), to changes in the current of the independent source $\Delta\bar{\imath}$, in active and reactive components ($\imath_{\rm p}$, $\imath_{\rm q}$).  Formally:
\begin{align}
        \left[\begin{array}{cc}\Delta \rho\\\Delta \omega\end{array}\right]&=
        \left[\begin{array}{cc}S_{\rho\imath_{\rm p}}&S_{\rho\imath_{\rm q}}\\
        S_{\omega\imath_{\rm p}}&S_{\omega\imath_{\rm q}}\end{array}\right]
        \left[\begin{array}{cc}\Delta \imath_{\rm p}\\\Delta \imath_{\rm q}\end{array}\right] , \\
        \Delta\underline{\eta}'&=
        \ull{S}'\, \Delta \underline{\imath}_{\rm pq} \, . \label{eq:predictorone}       
\end{align}   

\subsubsection{Second-order strength}
Sensitivity of the second-order dynamic of the voltage vector, represented through the second-order CF $\Delta\bar{\eta}''$, in real and imaginary parts ($\sigma$, $\gamma$), to changes in the current of the independent source $\Delta\bar{\imath}$, in active and reactive components ($\imath_{\rm p}$, $\imath_{\rm q}$).  Formally:
\begin{align}
    \left[\begin{array}{cc}\Delta \sigma\\\Delta \gamma\end{array}\right]&=
    \left[\begin{array}{cc}S_{\sigma\imath_{\rm p}}&S_{\sigma\imath_{\rm q}}\\
    S_{\gamma\imath_{\rm p}}&S_{\gamma\imath_{\rm q}}\end{array}\right]
    \left[\begin{array}{cc}\Delta \imath_{\rm p}\\\Delta \imath_{\rm q}\end{array}\right] , \\
    \Delta\underline{\eta}''&=
    \ull{S}''\, \Delta \underline{\imath}_{\rm pq} \, . \label{eq:sdef22}   
\end{align} 

According to the proposed general formulation, voltage strength at a single bus is quantified through twelve individual indicators, four per strength order.  In practice, some symmetries in actual power system models cause some indicators to be equal, such as $|S_{v\imath_{\rm q}}|\approx |S_{\theta\imath_{\rm p}}|$ and $|S_{v\imath_{\rm p}}|\approx |S_{\theta\imath_{\rm q}}|$. Nevertheless, in the most general case, each bus has a set of twelve individual strength indicators. A remarkable feature of our formulation is the clear domain of the indicators, facilitating their understanding by making their physical interpretation explicit. For instance, $S_{v\imath_{\rm q}}$ is the zero-order strength metric for the voltage's magnitude with respect to changes in reactive current injections. $S_{\gamma\imath_{\rm p}}$ is the second-order strength metric for the imaginary part of the second-order CF with respect to changes in the active current injections, and so forth. 
Identifying specific correlations between individual metrics and potential stability problems is out of the scope of the paper. However, we can safely anticipate that two of them, $S_{v\imath_{\rm q}}$ and $S_{\gamma\imath_{\rm p}}$, are closely related to the study of voltage stability and frequency stability, because of their close connection with conventional metrics, namely the SCL and nodal inertia, respectively.

\subsection{Assumptions for the Evaluation of Strength Indicators}
\label{subsec:strength_assumptions}

To estimate the set of indicators presented in \eqref{eq:sdef00}-\eqref{eq:sdef22}, we propose an analytical approach based on a dynamic model of the system.  The resulting indicators are therefore a function of the variables and parameters of the model.  As any mathematical model representing a physical system, it is based on a set of working assumptions.  The most important assumptions are as follows.
\begin{itemize}
\item Three-phase systems in balanced conditions.  This allows representing voltages and currents using dynamic Clarke vectors, and the objects introduced in the Addendum provided with this paper.  In particular, the reference used for voltages and currents is an angle rotating at the fundamental frequency $\omega_0$.
\item Fast electromagnetic dynamics are treated as algebraic constraints.  Consequently, a constant admittance matrix models the transmission network.  Consistently, the effect of fast-acting controllers on the same time scale is considered instantaneous.
\end{itemize}

This model is the reference of exactness in this work, and is the starting point of the derivation presented below.  Note that the formulation is compatible with data-driven approaches, i.e., the proposed indicators can also be calculated based on simulation signals and/or actual measurements. In this paper, we follow an analytical approach. A discussion on data-driven techniques is beyond the scope of this work.

\subsection{Derivation}
\label{subsec:derivation}

The starting point is the dynamic model in the standard form of a set of Differential-Algebraic Equations (DAEs).  The goal is to find an analytical expression for the strength indicators introduced above, in terms of the parameters $\bfg p$ and variables ($\bfg x(t)$ $\bfg y(t)$) of the system of DAEs.  Formally:
\begin{align}
    \Delta\bfg{\underline{v}}_{\rm v\theta}&=\szero(\bfg p,\bfg x(t),\bfg y(t))\Delta\bfg{\underline{\imath}}_{\rm pq}\,,\label{eq:dv_sought} \\
    \Delta\bfg{\underline{\eta}}'&=\sone(\bfg p,\bfg x(t),\bfg y(t))\Delta\bfg{\underline{\imath}}_{\rm pq}\,, \label{eq:deta_sought}\\
    \Delta\bfg{\underline{\eta}}''&=\stwo(\bfg p,\bfg x(t),\bfg y(t))\Delta\bfg{\underline{\imath}}_{\rm pq}\,,\label{eq:detaa_sought}
\end{align}
where $\Delta\bfg{\underline{\imath}}_{\rm pq}$ is a column vector containing the current injection of an independent current source at each bus (in $\rm pq$ coordinates relative to each bus), whereas $\Delta\bfg{\underline{v}}_{\rm v\theta}$, $\Delta\bfg{\underline{\eta}}'$, $\Delta\bfg{\underline{\eta}}''$ are column vectors containing the change due to $\Delta \bfg{\underline{\imath}}_{\rm pq}$ in the voltage, the first-order CF, and the second-order CF, at each bus, respectively. Therefore, $\szero$, $\sone$, and $\stwo$ are square matrices containing information on strength. In particular, the sought indicators are those on the diagonal, namely the bus voltage sensitivities to changes in current injection at the same bus. Each element of the diagonal is a smaller square matrix (2x2) containing the strength metrics as defined in \eqref{eq:sdef00}-\eqref{eq:sdef22}.
The off-diagonal elements of $\szero$, $\sone$, and $\stwo$ are in turn a measure of the sensitivity of a bus with respect to changes in the current injection at a different bus, and a byproduct of the formulation with relevant information that can be used in future work.

The dependence on time of $\bfg x(t)$ and $\bfg y(t)$ is at two specific times: right before ($-$) and right after ($+$) the current injection change introduced by $\Delta\bfg{\underline{\imath}}_{\rm pq}$.  Thus, \eqref{eq:dv_sought}-\eqref{eq:detaa_sought} can be expressed more precisely as follows:
\begin{align}
    \Delta\bfg{\underline{v}}_{\rm v\theta}&=\szero(\bfg p,\bfg x^{-},\bfg y^{-},\bfg x^{+},\bfg y^{+})\Delta\bfg{\underline{\imath}}_{\rm pq}\,,\label{eq:desired_form_1}\\
     \Delta\bfg{\underline{\eta}}'&=\sone(\bfg p,\bfg x^{-},\bfg y^{-},\bfg x^{+},\bfg y^{+})\Delta\bfg{\underline{\imath}}_{\rm pq}\,,\label{eq:desired_form_2}\\
    \Delta\bfg{\underline{\eta}}''&=\stwo(\bfg p,\bfg x^{-},\bfg y^{-},\bfg x^{+},\bfg y^{+})\Delta\bfg{\underline{\imath}}_{\rm pq}\,,\label{eq:desired_form_3}
\end{align}

The dependence on pre-disturbance values $\bfg x^{-}$ and $\bfg y^{-}$ highlights that the strength indicators are specific to a given operating condition.  This also implies that calculating their numerical values requires initializing the set of DAEs beforehand.  The strength metrics are also a function of post-disturbance values $\bfg{x^{+}, y^{+}}$, i.e., the resistance of the voltage to a perturbation depends not only on the system parameters and current operating condition, but also on the perturbation itself.  This is a very important aspect of the formulation, which retains the nonlinearity of the system equations, and is a fundamental difference with respect to calculating small signal sensitivities.  Obtaining $\bfg x^{+}$ is straightforward as states do not jump on discrete events, i.e.,  $\bfg x^{+}=\bfg x^{-}$.  

% In turn, obtaining the post-disturbance value of algebraic variables requires more work.  A dedicated discussion on the need of $\bfg y^{+}$ is presented in the second part of the series.

The derivation starts by considering the set of DAEs of the system.  Without lack of generality, we assume it is in a current injection form, i.e., each shunt-connected device interfaces with the transmission network through its current injection at the connection bus \cite{milanoscripting}.  The algebraic equations of the network voltages and current injections are:
\begin{equation}\label{eq:startmat}
    \vvec=\Zmat\, \Ivec\,,
\end{equation}
where $\vvec$ and $\Ivec$ are column vectors containing the voltage and net current injected by shunt-connected devices at each bus, respectively.  $\Zmat$ is the impedance matrix, i.e., the inverse of the admittance matrix of the network.

We add the current injected by the fictitious independent source at each bus $\ivec$ to \eqref{eq:startmat}:
\begin{equation}\label{eq:startmat2}
    \vvec=\Zmat\, \Ivec + \Zmat\,\ivec\,.
\end{equation}

At this point, we organize the remainder of the derivation into six steps:
\begin{enumerate}
\item Use \eqref{eq:startmat2} to find expressions for $\vvec$, $\dot{\vvec}$ and $\ddot{\vvec}$ in terms of $\Ivec$, $\dIvec$, $\ddIvec$, and $\ivec$.
\item Apply the $\Delta$ operator to the equations found in the previous step to get expressions for $\dv$, $\Delta\dot{\vvec}$, and $\Delta\ddot{\vvec}$ in terms of $\Delta\Ivec$, $\Delta\dIvec$, $\Delta\ddIvec$, and $\di$.  Note that at this point, continuing the derivation requires replacing $\Delta\Ivec$, $\Delta\dIvec$, and $\Delta\ddIvec$ depending on device models.
\item For a single generic device, write expressions for $\Delta\bar{\imath}_{\rm dev}$, $\Delta\dot{\bar{\imath}}_{\rm dev}$, and $\Delta\ddot{\bar{\imath}}_{\rm dev}$, as a function of $\Delta\bar{v}$, $\Delta\dot{\bar{v}}$, and $\Delta\ddot{\bar{v}}$.  
\item Combine the equations of steps 2 and 3 to solve for $\dv$, $\Delta\dot{\vvec}$, and $\Delta\ddot{\vvec}$.  The solutions depend on $\ivec$, the system's parameters, and the system's variables.  
\item Apply the transformations required to express $\dv$, $\Delta\dot{\vvec}$ and $\Delta\ddot{\vvec}$ in the desired form, i.e., as $\dvs$, $\detas$ and $\detaas$, and also $\Delta\ivec$ as $\dis$.
\item The derivation concludes by combining the results of steps 4 and 5 to get the sought expressions for the three orders of strength metrics, i.e., $\szero$, $\sone$, and $\stwo$.
\end{enumerate}

%with specific expressions depending on the device models composing the grid.  Some of them for common power system devices are presented in \ref{sec:devmodels}, and analytical examples using them to find the final expressions for the strength indicators, i.e., $\szero$, $\sone$ and $\stwo$ are presented in \ref{sec:anal_exam}.  
\subsubsection{Step 1}

Equation \eqref{eq:startmat2} is already in the desired form at this stage:
\begin{equation}
    \label{eq:step1_v}
    \vvec=\Zmat\, \Ivec + \Zmat\,\ivec\,.
\end{equation}

By applying the time derivative to \eqref{eq:step1_v}, we find the expressions for $\dot{\vvec}$ and $\ddot{\vvec}$:
\begin{align}
    \dot{\vvec}&=\Zmat\, \dIvec\,,\label{eq:step1_vdot}\\
    \ddot{\vvec}&=\Zmat\, \ddIvec\,,\label{eq:step1_vddot}
\end{align}
where $\dot{\ivec}=\ddot{\ivec}=0$ as $\ivec$ is a fictitious independent source, acting as a step-like input.

\subsubsection{Step 2}

We apply the $\Delta$ operator to \eqref{eq:step1_v}-\eqref{eq:step1_vddot}:
\begin{align}
    \dv&=\Zmat\, \Delta\Ivec + \Zmat\,\Delta\ivec\,,\label{eq:step2_v}\\
    \Delta\dot{\vvec}&=\Zmat\,\Delta\dIvec\,,\label{eq:step2_vdot}\\
    \Delta\ddot{\vvec}&=\Zmat\,\Delta\ddIvec\,\label{eq:step2_vddot}.
\end{align}

\subsubsection{Step 3}

\begin{figure}[hbtp]
    \centering
    \includegraphics[width=0.35\textwidth]{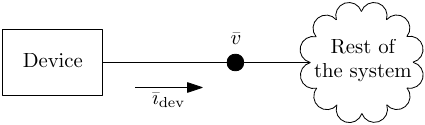}
    \caption{Generic device interface with the rest of the system.}
    \label{fig:device}
\end{figure}

Consider a single device shunt-connected at a generic bus as illustrated in Fig.~\ref{fig:device}.

In a current injection form, the device can be seen as an input-output block, where the input is the terminal voltage, $\bar{v}$, and the output is the current injected at the bus, $\bar{\imath}_{\rm dev}$. This consideration allows us to express the change in the current as a linear function of the change in the voltage:
\begin{equation}
  \label{eq:didevform}
  \Delta\ul{\imath}_{\rm dev}=\ull{a}\,\Delta\ul{v}\,,
\end{equation}
where $\ull{a}$ depends on variables and parameters of the specific model of the device.  Note that we have replaced the use of complex numbers with an equivalent representation with vectors and matrices (see the Addendum provided with this paper for a detailed explanation on the mathematical objects used and their notation).  Consequently, $\ull{a}$ has four degrees of freedom instead of two, making \eqref{eq:didevform} more general.  

%{\color{red} IgP: more things can be said regarding this point...  For example, using the matricial form is more general because you don't restrict your formulation to models that are a complex function of $\bar{v}$.  There should be a more formal way to say this...}

Similarly, we can assume that the time derivatives of the current are ultimately a function of the voltage and its time derivatives. This allows us to express $\Delta\dot{\ul{\imath}}_{\rm dev}$ and $\Delta\ddot{\ul{\imath}}_{\rm dev}$ as follows:
\begin{align}
    \Delta\dot{\ul{\imath}}_{\rm dev}&=\ull{a}'\, \Delta\ul{v}+\ull{b}'\, \Delta\dot{\ul{v}}\,,\\
    \Delta\ddot{\ul{\imath}}_{\rm dev}&=\ull{a}''\, \Delta\ul{v}+\ull{b}''\, \Delta\dot{\ul{v}}+\ull{c}''\, \Delta\ddot{\ul{v}}\,.
\end{align}
%{\color{red}IgP: What's best notation for $a$, $b$, $c$? Matricial form?? better selection of symbols...}

Expressions for $\ull{a}$, $\ull{a}'$, $\ull{a}''$, $\ull{b}'$, $\ull{b}''$, and $\ull{c}''$ specific to relevant power system device models are presented in Section \ref{sec:devmodels}.

\subsubsection{Step 4}

The results of the previous step can be written for all buses as follows:
\begin{align}
    \Delta\vIvec&=\ull{\bfg A}\,  \vdv\,,\label{eq:step4_v}\\
    \Delta\vdIvec&=\ull{\bfg A}'\,  \vdv+\ull{\bfg B}'\,  \vdvdot\,,\label{eq:step4_vdot}\\
    \Delta\vddIvec&=\ull{\bfg A}''\,  \vdv+\ull{\bfg B}''\,  \vdvdot+\ull{\bfg C}''\,  \vdvddot\,.\label{eq:step4_vddot}
\end{align}

Replacing \eqref{eq:step4_v} in \eqref{eq:step2_v}:
\begin{align}
    \vdv&=\mZmat\,  \ull{\bfg A}\,  \vdv + \mZmat\, \Delta\vivec\,,\\
    \Leftrightarrow \vdv&=\left(\Iden - \mZmat\,  \ull{\bfg A}\right)^{-1}\, \mZmat\, \Delta\vivec\,,
\end{align}
where $\Iden$ is the identity matrix. Denote:
\begin{equation}
\mZmateq:=\left(\Iden - \mZmat\,  \ull{\bfg A}\right)^{-1}\, \mZmat\,,\\
\end{equation}
then:
\begin{equation}
    \label{eq:step4_vres}
    \vdv=\mZmateq\, \Delta\vivec\,.  
\end{equation}

Equation \eqref{eq:step4_vres} is the sought expression at this step for $\vdv$.  We continue replacing \eqref{eq:step4_vdot} in \eqref{eq:step2_vdot}:
\begin{align}
    \vdvdot&=\mZmat\, \left(\ull{\bfg A}'\,  \vdv+\ull{\bfg B}'\,  \vdvdot\right)\,,\\
    \Leftrightarrow\vdvdot&= \left(\Iden - \mZmat\,  \ull{\bfg B}'\right)^{-1}\, \mZmat\, \ull{\bfg A}'\, \vdv\,. 
\end{align}

Using the previous result in \eqref{eq:step4_vres}:
\begin{equation}
    \vdvdot= \left(\Iden - \mZmat\,  \ull{\bfg B}'\right)^{-1}\, \mZmat\, \ull{\bfg A}'\, \mZmateq\, \Delta\vivec\,.
\end{equation}

Denote:
\begin{equation}
    \mZmateqp:=\left(\Iden - \mZmat\,  \ull{\bfg B}'\right)^{-1}\, \mZmat\, \ull{\bfg A}'\, \mZmateq\,,
\end{equation}
then:
\begin{equation}
    \label{eq:step4_vdotres}
    \vdvdot=\mZmateqp\, \Delta\vivec\,.  
\end{equation}

Equation \eqref{eq:step4_vdotres} is the sought expression at this step for $\vdvdot$.  Repeating the procedure for $\vdvddot$:
\begin{align}
    \vdvddot&=\mZmat\, \left(\ull{\bfg A}''\,  \vdv+\ull{\bfg B}''\,  \vdvdot+\ull{\bfg C}''\,  \vdvddot\right)\,,\\
    \Leftrightarrow\vdvddot&= \left(\Iden - \mZmat\,  \ull{\bfg C}''\right)^{-1}\,\left(\mZmat\, \ull{\bfg A}''\, \vdv+\mZmat\, \ull{\bfg B}''\, \vdvdot\,\right) .
\end{align}

Using the previous results in \eqref{eq:step4_vres} and \eqref{eq:step4_vdotres}:
\begin{equation}
    \vdvddot=\left(\Iden - \mZmat\,  \ull{\bfg C}''\right)^{-1}\,\left( \mZmat\, \ull{\bfg A}''\, \mZmateq+\mZmat\, \ull{\bfg B}''\, \mZmateqp\right)\, \Delta\vivec\,.
\end{equation}
Denote:
\begin{equation}
    \mZmateqpp:=\left(\Iden - \mZmat\,  \ull{\bfg C}''\right)^{-1}\, \left(\mZmat\, \ull{\bfg A}''\, \mZmateq+\mZmat\, \ull{\bfg B}''\, \mZmateqp\right)\,,
\end{equation}
then:
\begin{equation}
    \label{eq:step4_vddotres}
    \vdvddot=\mZmateqpp\, \Delta\vivec\,.  
\end{equation}

Equation \eqref{eq:step4_vddotres} is the sought expression for $\vdvddot$.

\subsubsection{Step 5}

We take the resulting expressions of the previous step and transform them into the desired form proposed in \ref{subsec:strength_proposal}.  We start with the current, which we require to transform from $\Delta\vivec$ to $\dis$.  The former is in global $\rm dq$ coordinates with respect to a common rotating reference frame at $\omega_0$, and the latter is with respect to the angle of the voltage corresponding to each bus.  This transformation makes the d-axis and q-axis currents the active and reactive components, respectively, which motivates the subscript $\rm pq$.  

We start by applying the inverse Park transform to $\vivec$:
\begin{equation}
   \label{eq:step5_1}
    \Delta\vivec = \Delta\left\{\ull{\bfg e}^{\jmath\,\theta}\, \vivec_{\rm pq}\right\}\,,
\end{equation}
where $\ull{\bfg e}^{\jmath\,\theta}$ is a diagonal matrix whose k-th component is:
\begin{equation}
    \ull{e}^{\jmath\,\theta}_k=\begin{bmatrix}
    \cos(\theta_k)&-\sin(\theta_k)\\
    \sin(\theta_k)&\cos(\theta_k)
    \end{bmatrix}\,,
\end{equation}
and $\theta_k$ is the angle of the voltage vector at bus $k$ with respect to the global $\rm dq$ reference frame rotating at $\omega_0$.

Next, we apply the definition of the $\Delta$ operator (equation \eqref{eq:def_delta_simp}.  Hence, \eqref{eq:step5_1} becomes:
\begin{equation}
    \Delta\vivec = \ull{\bfg e}^{\jmath\,\theta^{+}}\, \vivec^{+}_{\rm pq}-\ull{\bfg e}^{\jmath\,\theta^{-}}\, \vivec^{-}_{\rm pq}\,.
\end{equation}
Note that $\vivec^{-}_{\rm pq}=0$ by definition, and thus $\vivec^{+}_{\rm pq}=\Delta\vivec$.  Therefore:
\begin{equation}
    \label{eq:step5_i}
    \Delta\vivec = \ull{\bfg e}^{\jmath\,\theta^{+}}\, \Delta\vivec_{\rm pq}\,.
\end{equation}

Equation \eqref{eq:step5_i} is the sought transformation for the current.  We continue by deriving the expression required to transform $\Delta\ul{\bfg v}$ into $\dvs$.  Recalling the alternative form of identity 1 (equation \eqref{eq:identity1_alt} and using a matrix representation:
\begin{equation}
    \label{eq:step5_v}
    \Delta\ul{\bfg v}=\widetilde{\bfg v}\, \widetilde{\ull{\bfg e}^{\jmath\,\theta}}\, \dvs \, .
\end{equation}

Equation \eqref{eq:step5_v} is the sought transformation for the voltage.  Next, we find the transformation from $\Delta\dot{\ul{\bfg v}}$ to $\detas$.  To do so, we use the complex frequency property of acting as a time derivative operator as presented in equation \eqref{eq:cf_prop1_5}:
\begin{align}
    \Delta\dot{\ul{\bfg v}}&=\Delta\left\{\ull{\bfg v}\, \ul{\bfg \eta}'\right\}\,,\\
    \Delta\dot{\ul{\bfg v}}&=\uletavecam\, \vdv+\widetilde{\ull{\bfg v}}\, \detas\,.
\end{align}

Therefore:
\begin{equation}
    \label{eq:step5_eta}
    \detas=\widetilde{\ull{\bfg v}}^{-1}\, \Delta\dot{\ul{\bfg v}}-\widetilde{\ull{\bfg v}}^{-1}\, \uletavecam \, \vdv \, .
\end{equation}

Equation \eqref{eq:step5_eta} is the sought transformation for the first-order complex frequency.  Finally, the transformation for the second-order complex frequency is found following a similar procedure:
\begin{equation}
    \label{eq:step5_etaa}
    \detaas=\widetilde{\ull{\bfg v}}^{-1}\, \Delta\ddot{\ul{\bfg v}}-\widetilde{\ull{\bfg v}}^{-1}\, \uletaavecam \, \vdv\,.
\end{equation}

\subsubsection{Step 6}

We combine the results of steps 4 and step 5 to get the final expressions for $\szero$, $\sone$, and $\stwo$.  First, we replace \eqref{eq:step5_i} and \eqref{eq:step5_v} into \eqref{eq:step4_vres}:
\begin{equation}
    \dvs=\widetilde{\bfg v}^{-1}\, (\widetilde{\ull{\bfg e}^{\jmath\,\theta}})^{-1}\, \mZmateq\, \ull{\bfg e}^{\jmath\,\theta^{+}}\, \Delta\vivec_{\rm pq}\,.
\end{equation}

Therefore, the sought expression for the zero-order is:
\begin{equation}
\label{eq:sought_szero}
\boxed{
    \szero = \widetilde{\bfg v}^{-1}\, (\widetilde{\ull{\bfg e}^{\jmath\,\theta}})^{-1}\, \mZmateq\, \ull{\bfg e}^{\jmath\,\theta^{+}}
}
\end{equation}

Next, we replace \eqref{eq:step4_vres}, \eqref{eq:step4_vdotres} and \eqref{eq:step5_i} into \eqref{eq:step5_eta}:
\begin{equation}
    \detas=\widetilde{\ull{\bfg v}}^{-1}\, \left(\mZmateqp- \uletavecam \, \mZmateq\right)\, \ull{\bfg e}^{\jmath\,\theta^{+}}\, \Delta\vivec_{\rm pq}\,.
\end{equation}

Therefore, the sought expression for the first-order is:
\begin{equation}\label{eq:sought_sone}
\boxed{
    \sone = \widetilde{\ull{\bfg v}}^{-1}\, \left(\mZmateqp- \uletavecam \, \mZmateq\right)\, \ull{\bfg e}^{\jmath\,\theta^{+}}
}
\end{equation}

We continue by replacing \eqref{eq:step4_vres}, \eqref{eq:step4_vddotres} and \eqref{eq:step5_i} into \eqref{eq:step5_etaa}:
\begin{equation}
    \detaas=\widetilde{\ull{\bfg v}}^{-1}\, \left(\mZmateqpp- \uletaavecam \, \mZmateq\right)\, \ull{\bfg e}^{\jmath\,\theta^{+}}\, \Delta\vivec_{\rm pq}\,.
\end{equation}

Hence, the sought expression for the second-order is:
\begin{equation}\label{eq:sought_stwo}
\boxed{
    \stwo = \widetilde{\ull{\bfg v}}^{-1}\, \left(\mZmateqpp- \uletaavecam \, \mZmateq\right)\, \ull{\bfg e}^{\jmath\,\theta^{+}}
}
\end{equation}
where:
\begin{align}
    \mZmateq&=\left(\Iden - \mZmat\,  \ull{\bfg A}\right)^{-1}\, \mZmat\,,\\
    \mZmateqp&=\left(\Iden - \mZmat\,  \ull{\bfg B}'\right)^{-1}\, \mZmat\, \ull{\bfg A}'\, \mZmateq\,,\\
    \mZmateqpp&=\left(\Iden - \mZmat\,  \ull{\bfg C}''\right)^{-1}\, \mZmat\, \left( \ull{\bfg A}''\, \mZmateq+\ull{\bfg B}''\, \mZmateqp\right)\,.
\end{align}

\begin{figure}[hbtp]
    \centering
    \includegraphics[width=0.575\linewidth]{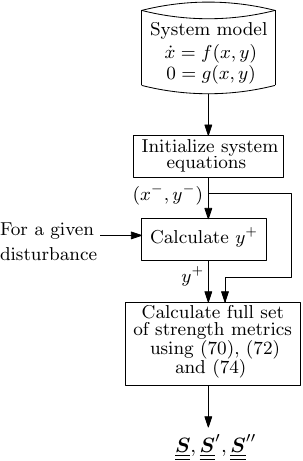}
    \caption{High-level flowchart of the process for calculating the full set of strength metrics.}
    \label{fig:flowchart}
\end{figure}

The resulting strength metrics depend on the device models through $\ull{\bfg A}$, $\ull{\bfg A}'$, $\ull{\bfg A}''$, $\ull{\bfg B}'$, $\ull{\bfg B}''$ and $\ull{\bfg C}''$ matrices.  Consequently, evaluating strength as a function of the parameters and variables of a specific system requires knowledge of the devices composing the grid.  The problem thus reduces to find expressions for $\ull{a}$, $\ull{a}'$, $\ull{a}''$, $\ull{b}'$, $\ull{b}''$ and $\ull{c}''$ for specific devices.  We address this task in the next section.  Figure \ref{fig:flowchart} shows a flowchart of the process required to apply (\ref{eq:sought_szero}), (\ref{eq:sought_sone}) and (\ref{eq:sought_stwo}) to calculate the full set of strength metrics.

\section{Device models}
\label{sec:devmodels}

In this section, we work with a variety of basic power system device models that are relevant to system strength, e.g., synchronous machines, inverter-based resources, and basic load models.  For each case, the goal is to find analytical expressions for $\ull{a}$, $\ull{a}'$, $\ull{a}''$, $\ull{b}'$, $\ull{b}''$, and $\ull{c}''$, which can be replaced into the results of the previous section to evaluate strength.

To do so, we follow a systematic methodology divided into two steps.  First, starting from the set of DAEs composing the dynamic model of the device, find expressions for its current injection at terminals and its time derivatives, i.e., $\videv$, $\videvdot$, and $\videvddot$.  These expressions must be a function only of model inputs, states, the terminal voltage, and/or its time derivatives.  Second, apply the $\Delta$ operator to the equations found.  Finally, the sought expressions for $\ull{a}$, $\ull{a}'$, $\ull{a}''$, $\ull{b}'$, $\ull{b}''$, and $\ull{c}''$ can be identified from the results of the second step.

\subsection{Synchronous machines}
\label{subsec:syn2}

Consider the classical model of synchronous machines \cite{milanoscripting}, where the symbols have the usual meanings:
\begin{align}
    \dot{\delta}&=\Omega_{\rm b}(\omega_{\rm r}-1)\,,\\
    M\dot{\omega}_{\rm r}&=p_{\rm m}-p_{\rm e}-D(\omega_{\rm r}-1)\,,
\end{align}
along with the algebraic equations:
\begin{align}
    0&=(r_{\rm a}+\jmath\, x_{1{\rm d}})\,\idev-\bar{E}+\bar{v}\,,\\
    0&=\bar{E}\exp(-\jmath\,(\delta-\pi/2))-\jmath\, e_{1\rm q}\,,\\
    0&=\Re\{\bar{E}\,\idev^{*}\}-p_{\rm e}\,,\label{eq:pe_syn2}
\end{align}
where $p_{\rm m}$ and $e_{1\rm q}$ are the inputs of the model, normally states of the machine controllers.

From the set of DAEs of the machine's model, the following expressions are found:
\begin{align}
    \idev&=(r_{\rm a}+\jmath\, x_{1{\rm d}})^{-1}\left(\bar{E}-\bar{v}\right)\,,\label{eq:syn2_idev}\\
    \idevdot&=(r_{\rm a}+\jmath\, x_{1{\rm d}})^{-1}\left(\dot{\bar{E}}-\dot{\bar{v}}\right)\,,\label{eq:syn2_idevdot}\\
    \idevdot&=(r_{\rm a}+\jmath\, x_{1{\rm d}})^{-1}\left(\ddot{\bar{E}}-\ddot{\bar{v}}\right)\,,\label{eq:syn2_idevddot}
\end{align}
where:
\begin{align}
    \bar{E}&=e_{1\rm q}\exp(\jmath\,\delta)\,,\label{eq:e_syn2_0}\\
    \dot{\bar{E}}&=\bar{E}\left(\frac{\dot{e}_{1\rm q}}{e_{1\rm q}}+\jmath\,\dot{\delta}\right)\,,\label{eq:de_syn2_0}\\
    \ddot{\bar{E}}&=\bar{E}\left(\left(\frac{\dot{e}_{1\rm q}}{e_{1\rm q}}+\jmath\,\dot{\delta}\right)^{2}+\frac{\ddot{e}_{1\rm q}e_{1\rm q}-\dot{e}^{2}_{1\rm q}}{{e_{1\rm q}}^{2}}+\jmath\,\ddot{\delta}\right)\,.\label{eq:dde_syn2_0}
\end{align}

Considering a constant $e_{1\rm q}$ and the differential equations of the model, equations \eqref{eq:e_syn2_0}-\eqref{eq:dde_syn2_0} can be rewritten as:
%\small
\begin{align}
    \bar{E}&=e_{1\rm q}\exp(\jmath\,\delta)\label{eq:e_syn2_1}\\
    \dot{\bar{E}}&=\jmath\,\Omega_{\rm b}\bar{E}(\omega_{\rm r}-1)\label{eq:de_syn2_1}\\
    \ddot{\bar{E}}&=\bar{E}(-\Omega_{\rm b}^{2}(\omega_{\rm r}-1)^{2}+\jmath\,\Omega_{\rm b}\dot{\omega}_{\rm r})\\
    &=\bar{E}(-\Omega_{\rm b}^{2}(\omega_{\rm r}-1)^{2}+\jmath\,\frac{\Omega_{\rm b}}{M}\left(p_{\rm m}-p_{\rm e}-D(\omega_{\rm r}-1)\right)) \, .
    \label{eq:dde_syn2_1}
\end{align}

The $\Delta$ operator is applied to \eqref{eq:syn2_idev}:
\begin{equation}
    \Delta\idev=(r_{\rm a}+\jmath\, x_{1{\rm d}})^{-1}\left(\Delta\bar{E}-\Delta\bar{v}\right) \, ,
\end{equation}
where, using identity 1 (equation \eqref{eq:identity1_alt}:
\begin{equation}
    \Delta\bar{E} = \widetilde{e}_{1\rm q}\, \widetilde{{\rm e}^{\, \jmath\,\delta}} \left( \frac{\Delta e_{1\rm q}}{\widetilde{e}_{1\rm q}} + \jmath \, \frac{\tan(\Delta\delta/2)}{1/2} \right) \, .
\end{equation}

As $\delta$ is a state and $e_{1\rm q}$ a constant input, $\Delta\delta=\Delta e_{1\rm q}=0$.  Therefore:
\begin{equation}
    \Delta\bar{E}=0\,.\label{eq:DE_syn2}
\end{equation}

Finally, we obtain:
\begin{equation}
    \Delta\idev=-(r_{\rm a}+\jmath\, x_{1{\rm d}})^{-1}\Delta\bar{v}\,,
\end{equation}
from where $\ull{a}$ is identified:
\begin{equation}
\boxed{
    \ull{a}=-\begin{bmatrix}
    r_{\rm a} & -x_{1{\rm d}}\\
    x_{1{\rm d}} & r_{\rm a}
    \end{bmatrix}^{-1}
}
\end{equation}

Next, the $\Delta$ operator is applied to \eqref{eq:syn2_idevdot}:
\begin{equation}
    \Delta\idevdot=(r_{\rm a}+\jmath\, x_{1{\rm d}})^{-1}\left(\Delta\dot{\bar{E}}-\Delta\dot{\bar{v}}\right)\,,
\end{equation}
where:
\begin{equation}
    \Delta\dot{\bar{E}}=\jmath\,\Omega_{\rm b}\left(\Delta\bar{E}(\widetilde{\omega}_{\rm r}-1)+\bar{E}\Delta\omega_{\rm r}\right)\,.
\end{equation}
Again, as $\omega_{\rm r}$ is a state and using \eqref{eq:DE_syn2}:
\begin{equation}
    \Delta\dot{\bar{E}}=0\,.
\end{equation}
Hence:
\begin{equation}
    \Delta\idevdot=-(r_{\rm a}+\jmath\, x_{1{\rm d}})^{-1}\Delta\dot{\bar{v}}\,,
\end{equation}
from where $\ull{a}'$ and $\ull{b}'$ are identified:
\begin{equation}
\boxed{
    \ull{a}'=0\,;\qquad\ull{b}'=\ull{a} \, .
}
\end{equation}

Finally, we apply the $\Delta$ operator to \eqref{eq:syn2_idevddot}:
\begin{equation}\label{eq:didevddot_0}
    \Delta\idevddot=(r_{\rm a}+\jmath\, x_{1{\rm d}})^{-1}\left(\Delta\ddot{\bar{E}}-\Delta\ddot{\bar{v}}\right)\,,
\end{equation}
where:
\begin{equation}
    \Delta \ddot{\bar{E}}=\jmath\,\bar{E}\frac{\Omega_{\rm b}}{M}\left(\Delta p_{\rm m}-\Delta p_{\rm e}\right)\,.
\end{equation}

As $p_{\rm m}$ is considered a constant input:
\begin{equation}\label{eq:syn2_deddot_0}
    \Delta \ddot{\bar{E}}=-\jmath\,\bar{E}\frac{\Omega_{\rm b}}{M}\Delta p_{\rm e}\,.
\end{equation}

Note $p_{\rm e}$ is an algebraic variable of the machine model. To find the desired form formulated in \eqref{eq:step4_vddot}, we need to express $p_{\rm e}$ as a function of the current and terminal voltage.  To do so, note \eqref{eq:pe_syn2} can be equivalently rewritten as (see Addendum):
\begin{align}
    \begin{bmatrix}p_{\rm e}\\0
    \end{bmatrix}=\M{1}{0}{0}{0}\, \ull{E}^{*}\,  \videv\,.
\end{align}

Replacing in \eqref{eq:syn2_deddot_0} and using the previous result in \eqref{eq:syn2_idev} for $\Delta\videv$:
\begin{align}
    \Delta \ddot{\ul{E}}=-\jmath\,\ull{E}\, \frac{\Omega_{\rm b}}{M}\, \M{1}{0}{0}{0}\, \ull{E}^{*}\, \ull{a}\, \Delta\ul{v}\,.
\end{align}

The resulting equation for $\Delta\ddot{\ul{E}}$ is replaced in \eqref{eq:didevddot_0}:
\begin{align}
    \Delta\videvddot=\jmath\,\ull{a}\, \ull{E}\, \frac{\Omega_{\rm b}}{M}\, \M{1}{0}{0}{0}\, \ull{E}^{*}\, \ull{a}\, \Delta\ul{v}+\ull{a}\, \Delta\ddot{\ul{v}}\,,
\end{align}
from where $\ull{a}''$, $\ull{b}''$ and $\ull{c}''$ are identified:

\begin{equation}
\boxed{
    \begin{aligned}
    \ull{a}''&=\jmath\,\ull{a}\, \ull{E}\, \frac{\Omega_{\rm b}}{M}\M{1}{0}{0}{0}\, \ull{E}^{*}\, \ull{a}\,,\\
    \ull{b}''&=0\,;\qquad \ull{c}''=\ull{a} \, .
    \end{aligned}}
\end{equation}

\subsection{Grid-following (GFL) converters}
\label{subsec:gfl}

Consider a GFL converter model with active and reactive power control, and a droop frequency control.  In order to be consistent with the time scale of the rest of the models provided in this section, synchronization is considered ideal, i.e., the PLL dynamics are considered fast and neglected.\footnote{The only reason to neglect PLL dynamics is to match the slower time scale of the rest of the models provided in the paper, so that case studies in Section \ref{sec:study_case} are consistent. However, the framework accepts models with PLLs.} The set of DAEs describing such a model is shown below, where the symbols have the usual meanings:
\begin{align}
T\dot{\imath}_{\rm d}&=\imath_{\rm dref}-\imath_{\rm d}\,,\\
T\dot{\imath}_{\rm q}&=\imath_{\rm qref}-\imath_{\rm q}\,,\\
T_{\rm f}\dot{x}_{\rm p}&=\frac{1}{R}(\omega-\omega_{\rm ref})-x_{\rm p} \, ,
\end{align}
along with the algebraic equations:
\begin{align}
0&=\bar{s}_{\rm ref0}+x_{\rm p}-\bar{s}_{\rm ref}\,,\\
0&=\bar{v}_{\rm dq}\,\bar{\imath}_{\rm ref}^{*}-\bar{s}_{\rm ref}\,,\\
0&=\bar{v}_{\rm dq}-v\,,\\
0&=\bar{\imath}\,\etheta-\idev\,.
\end{align}

From the set of DAEs of the GFL model, the following expressions are found:
\begin{align}
\label{eq:I_gfl_3}
\idev&=\bar{\imath}\,\etheta\,;\quad \idevdot=\idev\left(\frac{\dot{\bar{\imath}}}{\bar{\imath}}+\jmath\,(\omega-\omega_0)\right)\,,\\ \nonumber
\idevddot&=\idev\left(\left(\frac{\dot{\bar{\imath}}}{\bar{\imath}}+\jmath\,(\omega-\omega_0)\right)^{2}+\jmath\,\dot{\omega}+\frac{\dot{\bar{\imath}}_{\rm ref}\,\bar{\imath}-\bar{\imath}_{\rm ref}\,\dot{\bar{\imath}}}{T\,\bar{\imath}^2}\right)\,,
\end{align}
where:
\begin{align}
\dot{\bar{\imath}}&=\frac{\bar{\imath}_{\rm ref}-\bar{\imath}}{T}\,,\\
\dot{\bar{\imath}}_{\rm ref}&=-K_{\rm p}\left(\frac{\dot{\omega}}{R}+\idev^{*}\,\dot{\bar{v}}+\idevdot^{*}\,\bar{v}\right)+K_{\rm i}\,(\bar{s}_{\rm ref}-\bar{s})\,.
\end{align}

Next, we apply the $\Delta$ operator to \eqref{eq:I_gfl_3} to find the expressions for $\Delta\idev$, $\Delta\idevdot$ and $\Delta\idevddot$:
\begin{align}\label{eq:di_gfl_0}
\Delta\idev&=\bar{\imath}\,\Delta \etheta.
\end{align}

Recalling property 5 and identity 1 (see \eqref{eq:prop5_alt} and \eqref{eq:identity1_alt}):
\begin{align}
\Delta\idev&=\bar{\imath}\,\widetilde{\etheta}\jmath\,\frac{\tan(\Delta\theta/2)}{1/2}\,,
\end{align}
which, in matrix form:
\begin{align}
\Delta\videv=\ull{\imath}\, 
\etilde\, \M{0}{0}{0}{1}
\widetilde{v}^{-1}\, (\etilde)^{-1}\, \Delta\ul{v}\,.
\end{align}

Therefore, $\ull{a}$ is identified:
\begin{equation}
\boxed{
\ull{a}=\ull{\imath}\, 
\etilde\, \M{0}{0}{0}{1}
\widetilde{v}^{-1}\, (\etilde)^{-1}\,.}
\end{equation}

In the case of $\Delta\idevdot$:
\begin{equation}\label{eq:didevdot_gfl0}
\begin{aligned}
    \Delta\idevdot=& \; \Delta \idev\left(\frac{\widetilde{\bar{\imath}}_{\rm ref}-\bar{\imath}}{T\,\bar{\imath}}+\jmath\,(\widetilde{\omega}-\omega_0)\right) \; +\\
    & \; \widetilde{\bar{\imath}}_{\rm dev}\left(\frac{\Delta\bar{\imath}_{\rm ref}}{T\,\bar{\imath}}+\jmath\,\Delta\omega\right)\,,
\end{aligned}
\end{equation}
where:
\begin{align}
    \Delta\omega&=\M{0}{1}{0}{0}\, 
    \widetilde{\ull{v}}^{-1}\, \left(\Delta\dot{\ul{v}}-\widetilde{\ull{\eta}}'\, \Delta\ul{v}\right)\,,\label{eq:dw_gfl}\\
    \Delta\ul{\imath_{\rm ref}}&=
   \M{-1}{0}{0}{1}
    \, \widetilde{\ull{v}}^{-1}\, \widetilde{\ull{\imath_{\rm ref}}}^{*}\, 
\M{1}{0}{0}{0}
    \, (\etilde)^{-1}\, 
    \Delta\ul{v}\,.\label{eq:diref_gfl}
\end{align}

Note the first term of \eqref{eq:didevdot_gfl0} and $\widetilde{\ull{\eta}}'$ are negligible compared to the other terms.  Adopting this simplification and combining \eqref{eq:didevdot_gfl0} to \eqref{eq:diref_gfl} leads to the sought expression for $\Delta\videvdot$:
\begin{equation}\label{eq:gfl_firstorder}
\begin{aligned}
    &\Delta\videvdot=\ull{\imath}\, 
\etilde\, 
\M{0}{0}{0}{1}
\, \widetilde{\ull{v}}^{-1}\, \Delta\ul{\dot{v}}\,+\\
&\frac{1}{T}\,\etilde\, 
\M{-1}{0}{0}{1}
\hat{v}^{-2}\, (\etilde)^{-1}\, \ull{s_{\rm ref}}
\, \M{1}{0}{0}{0}
\, (\etilde)^{-1}\, \Delta\ul{v}\,.
\end{aligned}
\end{equation}

Therefore, $\ull{a}'$ and $\ull{b}'$ are identified:
\begin{equation*}
\boxed{
\begin{aligned}
\ull{b}'&=
\ull{\imath}\, 
\etilde\, 
\M{0}{0}{0}{1}
\, \widetilde{\ull{v}}^{-1}\,,\\ 
\ull{a}'&=\frac{1}{T}\,\etilde\, 
\M{-1}{0}{0}{1}
\hat{v}^{-2}\, (\etilde)^{-1}\, 
\ull{s_{\rm ref}}\, 
\M{1}{0}{0}{0}
\, 
(\etilde)^{-1}\,.
\end{aligned}
}
\end{equation*}

Finally, a similar procedure is followed to find an expression for $\Delta\ddot{\bar{\imath}}_{\rm dev}$:
\begin{align}
&\Delta\videvddot=
\ull{\imath}\, 
\etilde\, 
\M{0}{0}{0}{1}\, 
\widetilde{\ull{v}}^{-1}\, \Delta\ddot{\ul{v}}\,+\\ \nonumber
&\frac{1}{T}\,\hat{v}^{-2}\etilde\, \left(\frac{1}{T_{\rm f}R}
\M{0}{1}{0}{0}-\ull{s_{\rm ref}}^{*}\, \M{1}{0}{0}{0}\right)
\, (\etilde)^{-1}\, \Delta\dot{\ul{v}}\,+\\ \nonumber
&\frac{1}{T^2}\,\widetilde{v}\,\etilde\, \hat{v}^{-4}\ull{\imath}^{-1}\, {\ull{s_{\rm ref}}^{*}}^{2}\, 
\M{1}{0}{0}{0}
\, (\etilde)^{-1}\, \Delta\ul{v}\,.
\end{align}
Therefore, $\ull{a}''$, $\ull{b}''$ and $\ull{c}''$ are found:
\begin{equation*}
\boxed{
\begin{aligned}
\ull{c}''&=\ull{b}'\,, \\ \ull{a}''&=\frac{1} {T^2}\,\widetilde{v}\,\etilde\, \hat{v}^{-4}\ull{\imath}^{-1}\, {\ull{s_{\rm ref}}^{*}}^{2}\, 
\M{1}{0}{0}{0}
\, (\etilde)^{-1}\,,\\
\ull{b}''&=\frac{1}{T}\,\hat{v}^{-2}\etilde\, \left(\frac{1}{T_{\rm f}R}
\M{0}{1}{0}{0}-\ull{s_{\rm ref}}^{*}\, \M{1}{0}{0}{0}\right)
\,(\etilde)^{-1} \, .
\end{aligned}
}
\end{equation*}
\subsection{Grid-forming (GFM) converters}
\label{subsec:gfm}
Consider a standard GFM converter connected to the grid through an output filter with series impedance $\bar{z}_{\rm f}$ and shunt conductance $\bar{y}_{\rm f}$. The control includes a power-frequency droop for synchronization and an integral block for voltage regulation. The set of DAEs of the model is shown below, where the symbols have the usual meanings:
\begin{align}
    \dot{\delta}&=\Omega_{\rm b}(\omega_{\rm gfm}-1)\,,\\
    \dot{e}&=K_{\rm i}(v_{\rm ref0}-v)\,,
\end{align}
along with the algebraic equations:
\begin{align}
    0&=m_{\rm p}(p_{\rm ref0}-p_{\rm e})+1-\omega_{\rm gfm}\\
    0&=\bar{e}-\bar{z}_{\rm f}(\idev+\bar{y}_{\rm f}\bar{v})-\bar{v}\,,\\
    0&=e\,\exp(\jmath\,\delta)-\bar{e}\,,\\
    0&=\bar{v}\,\idev^{*}-\bar{s}\,,\\
    0&=\Re\{\bar{s}\}-p_{\rm e}\,.
\end{align}

The following expressions are found from the set of DAEs:
\begin{align}
    \idev&=\bar{z}_{\rm f}^{-1}\,\bar{e}-\bar{z}_{\rm f}^{-1}(1+\bar{z}_{\rm f}\,\bar{y}_{\rm f})\,\bar{v}\,,\label{eq:gfm0}\\
    \idevdot&=\bar{z}_{\rm f}^{-1}\,\dot{\bar{e}}-\bar{z}_{\rm f}^{-1}(1+\bar{z}_{\rm f}\,\bar{y}_{\rm f})\,\dot{\bar{v}}\,,\\
    \idevddot&=\bar{z}_{\rm f}^{-1}\,\ddot{\bar{e}}-\bar{z}_{\rm f}^{-1}(1+\bar{z}_{\rm f}\,\bar{y}_{\rm f})\,\ddot{\bar{v}}\,,
\end{align}
where:
\begin{align}
\dot{\bar{e}}&=\bar{e}\left(\frac{\dot{e}}{e}+\jmath\,\dot{\delta}\right)\,,\quad \ddot{\bar{e}}=\bar{e}\left(\frac{\ddot{e}}{e}+2\jmath\frac{\dot{e}\,\dot{\delta}}{e}-\dot{\delta}^{2}+\jmath\,\ddot{\delta}\right)\,.\label{eq:gfm1}
\end{align}

By combining (\ref{eq:gfm0})-(\ref{eq:gfm1}) with the set of DAEs of the GFM, and following a similar procedure to that followed for the previous devices, the sought components are as follows.

Zero-order components:
\begin{equation}\label{eq:gfm_zeroorder}
\boxed{
\ull{a}=-\ull{z_{\rm f}}^{-1}(1+\ull{z_{\rm f}}\,\ull{y_{\rm f}})}
\end{equation}

First-order components:
\begin{equation}\nonumber
\boxed{
\begin{aligned}
\ull{b}'&=\ull{a}\,,\\
\ull{a}'&=-\ull{z_{\rm f}}^{-1}\ull{e}\left(\frac{K_{\rm i}}{e}\M{1}{0}{0}{0}(\etilde)^{-1}+\Omega_{\rm b}m_{\rm p}\M{0}{0}{1}{0}\ull{\imath_{\rm p}}\right)
\end{aligned}}
\end{equation}
where:
\begin{align}
    \ull{\imath_{\rm p}}=\widetilde{\ull{\imath_{\rm dev}}}^{*}+\widetilde{\ull{v}}\,\M{1}{0}{0}{-1}\,\ull{a}\,.
\end{align}

Second-order components:
\begin{equation}\label{eq:gfm_secondorder}
\boxed{
\begin{aligned}
\ull{c}''&=\ull{b}'\,,\\
\ull{b}''&=\ull{a}'\,,\\
\ull{a}''&=-\Omega_{\rm b}m_{\rm p}\,\ull{z_{\rm f}}^{-1}\ull{e}\M{0}{0}{1}{0}\widetilde{\ull{s}}\M{1}{0}{0}{-1}\widetilde{\ull{\imath_{\rm dev}}}^{-1}\ull{a}'
\end{aligned}}
\end{equation}

Differently from the GFL model, the GFM's zero-order component is constant and equal to the equivalent series admittance of the voltage source.  Key control parameters are the gains of the voltage control and the power-frequency droop constant.  A notable difference with respect to the GFL is that the droop affects both the first-order and the second-order components, i.e., those related to sudden changes in the frequency and the rate of change of the frequency. 

\subsection{Constant-impedance loads}
\label{subsec:zload}

Consider a standard constant shunt impedance load model:
\begin{equation}\label{eq:zload}
    0=\bar{v}+\bar{z}\,\idev\,.
\end{equation}
From \eqref{eq:zload}, the following expressions are found:
\begin{align}
\idev&=-\bar{z}^{-1}\,\bar{v}\,\Rightarrow\,\Delta\idev=-\bar{z}^{-1}\,\Delta\bar{v}\,,\\
\idevdot&=-\bar{z}^{-1}\,\dot{\bar{v}}\,\Rightarrow\,\Delta\idevdot=-\bar{z}^{-1}\,\Delta\dot{\bar{v}}\,,\\
\idevddot&=-\bar{z}^{-1}\,\ddot{\bar{v}}\,\Rightarrow\,\Delta\idevddot=-\bar{z}^{-1}\,\Delta\ddot{\bar{v}}\,.
\end{align}
Therefore:
\begin{equation}
\boxed{
\begin{aligned}   
\ull{a}&=\ull{b}'=\ull{c}''=-\ull{z}^{-1}\,,\\
\ull{a}'&=\ull{a}''=\ull{b}''=0\,.
\end{aligned}
}
\end{equation}

\section{Study cases}\label{sec:study_case}

The proposed framework is discussed based on the well-known IEEE 39-bus system, which has been modified to consider three scenarios: (i) classical machine model; (ii) substitution of part of the generation with GFL converters; and (iii) a system with 100\% IBR generation and a mix of GFL and GFM converters.
\subsection{Strength of conventional sycnhronous generators}
Synchronous machines are represented using the classical model described in Section \ref{subsec:syn2}.  Loads are considered constant impedances according to the model presented in Section \ref{subsec:zload}.  Standard system parameters and operating conditions are used and can be found in \cite{ieeepesTR18}.

\begin{figure}[hbtp]
    \centering
    \begin{subfigure}[]{0.48\linewidth}
        \caption{Indicator $S_{v\imath_{\rm q}}$.}
        \label{fig:s012}
        \includegraphics[width=\textwidth]{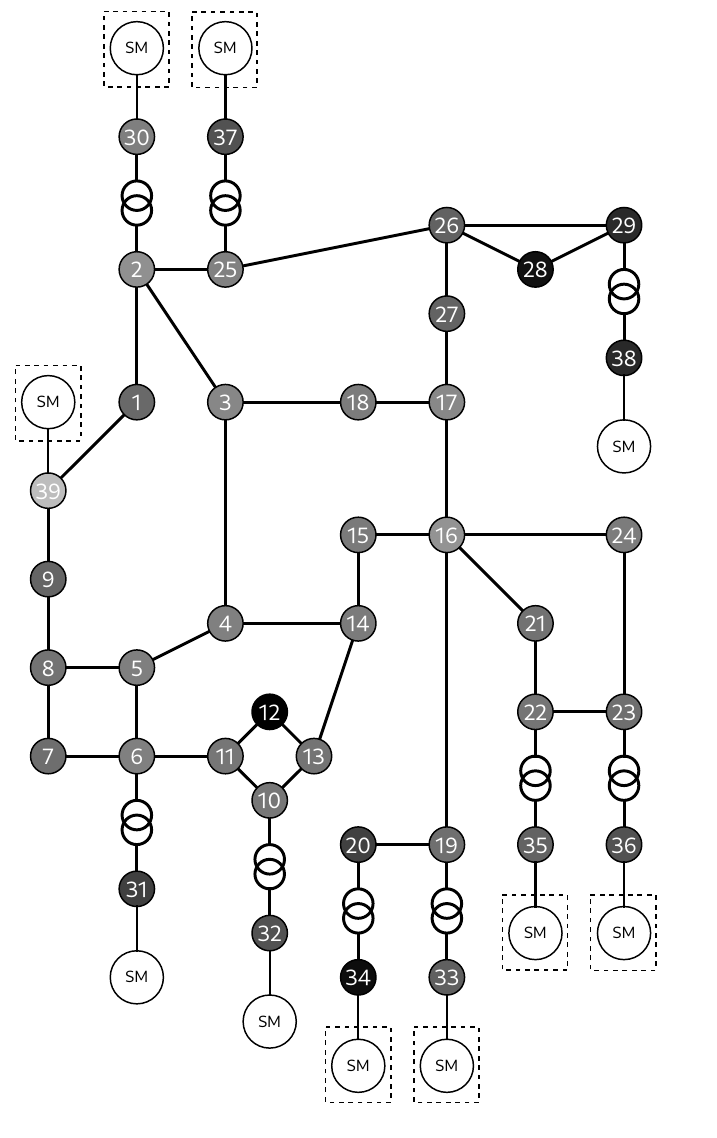}
    \end{subfigure}
    \hfill
    \begin{subfigure}[]{0.48\linewidth}
        \caption{Indicator $S_{\gamma\imath_{\rm p}}$.}
        \label{fig:s221}
        \includegraphics[width=\textwidth]{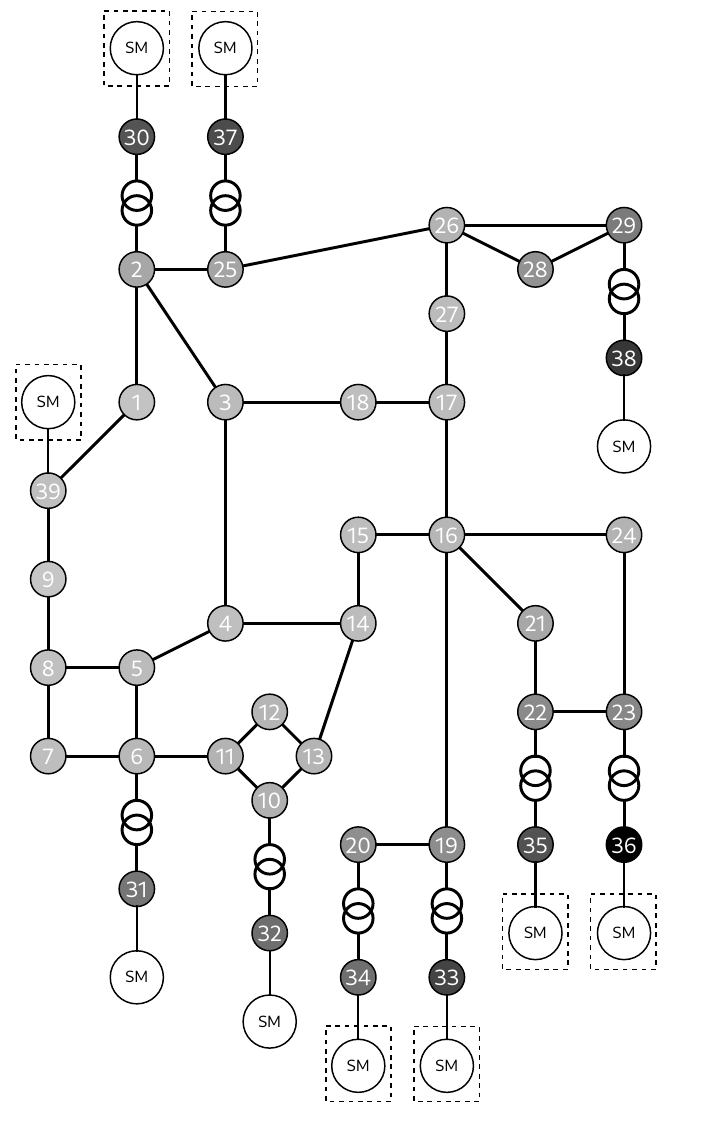}
    \end{subfigure}
    \hfill
    \includegraphics[width=0.9\linewidth]{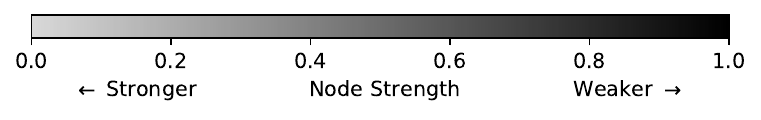}
    \caption{Normalized results of the system strength metrics considering a conventional power system.  (Note: generators marked with dashed rectangles are replaced by GLF converters in the second case study.)}
\label{fig:slds}
\end{figure}

The complete set of strength indicators $\szero$, $\sone$ and $\stwo$ is calculated at every bus using \eqref{eq:sought_szero}, \eqref{eq:sought_sone} and \eqref{eq:sought_stwo}.
Without lack of generality, the test perturbation used for this example is an active current sudden change of $\Delta \bar{\imath}_{\rm pq}=1+\jmath\,0$ pu.  The results for two representative strength indicators are presented in Fig.~\ref{fig:slds}.  The values are normalized to the maximum of all the indicators of the same order, and the absolute value is taken for easier comparison among different buses using a grayscale.  Fig.~\ref{fig:s012} shows the results for $S_{v\imath_{\rm q}}$, namely, the sensitivity of the magnitude of the voltage with respect to reactive current changes.  As expected, the results of this zero-order metric are driven by the topology of the network plus the internal impedance of SMs.  Nodes with a higher Thevenin equivalent are more sensitive, such as nodes 29, 28, and 38.  In turn, nodes in more meshed areas of the grid or closer to bigger SMs are stronger, such as nodes 02, 03, and 39.  The results for the opposite zero-order metric $S_{\theta\imath_{\rm p}}$, namely, the sensitivity of the angle of the voltage with respect to active current changes, is equal in magnitude to $S_{v\imath_{\rm q}}$.  This is a consequence of the device models used, particularly the classical model of SMs, which has a unique internal impedance for both coordinates.  

The other two zero-order metrics $S_{v\imath_{\rm p}}$ and $S_{\theta\imath_{\rm q}}$ are much lower than those discussed previously.  
Fig.~\ref{fig:s221} shows the results for $S_{\gamma\imath_{\rm p}}$, namely, the sensitivity of the RoCoF ($\gamma$ as defined in \eqref{eq:gammadef} with respect to active current changes.  In this case, buses closer to SMs terminals tend to be weaker than those further from them.  The rationale behind this result is that buses in `central' locations of the grid leverage the contribution of the inertia of all the SMs.  In turn, a SM terminal bus is mostly affected by the (relatively lower) inertia of the local machine.  In particular, bus 36 is the weakest bus, and coincides with the terminal bus of the machine of the lowest inertia in the system.  The rest of second-order metrics $S_{\gamma\imath_{\rm q}}$, $S_{\sigma\imath_{\rm p}}$ and $S_{\sigma\imath_{\rm q}}$ are negligible compared to $S_{\gamma\imath_{\rm p}}$.  

Finally, the four first-order metrics $S_{\rho\imath_{\rm p}}$, $S_{\rho\imath_{\rm q}}$, $S_{\omega\imath_{\rm p}}$, $S_{\omega\imath_{\rm q}}$ are null for this system.  This means that the first-order complex-frequency $\bar{\eta}'$ is infinitely strong, i.e., it does not jump after a sudden current change.  This is a consequence of the device models present in the system, particularly because of the classical model of SMs.  A system entirely composed of SMs makes the CF of the voltage at every node continuous.

\begin{figure}[hbtp]
    \centering
    \begin{subfigure}[]{0.49\linewidth}
        \caption{$\rho_{15}$ and its PRoC.}
        \label{fig:rho_15}
        \includegraphics[width=0.92\linewidth]{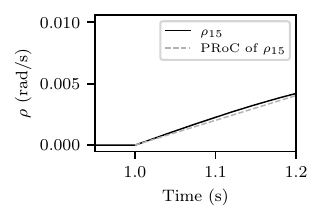}
    \end{subfigure}
    \hfill
    \begin{subfigure}[]{0.49\linewidth}
        \caption{$\omega_{15}$ and its PRoC.}
        \label{fig:om_15}
        \includegraphics[width=0.95\linewidth]{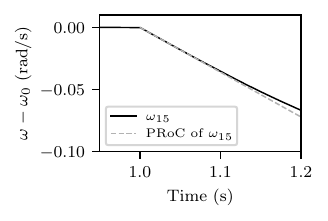}
    \end{subfigure}
    \caption{Trajectories of $\bar{\eta}'_{15}$ and their PRoC.}
    \label{fig:dynval}
\end{figure}

Having calculated the strength metrics of the system, a dynamic validation for a sample perturbation at bus 15 at simulation time $t=1$ s is done.  The initial deviations of $\Delta\bar{v}_{15}$, $\Delta\bar{\eta}'_{15}$ and $\Delta\bar{\eta}''_{15}$ predicted by the strength indicators for the given contingency are calculated using \eqref{eq:predictor0}, \eqref{eq:predictorone} and \eqref{eq:sdef22}, and compared with the actual trajectories of these variables observed after a time-domain simulation (TDS).  While $v(t)$ and $\theta(t)$ are directly available as they are part of the set of system DAEs, the exact trajectories of $\rho(t)$ and $\omega(t)$ are calculated using the Complex Frequency Divider Formula \cite{cfd}.  The results for these four variables are presented in Table \ref{tab:dynval}, which verifies the exactness of the formulation.  Regarding $\sigma(t)$ and $\gamma(t)$, their trajectories are unfortunately not available directly from the TDS.  However, as they are approximately equal to $\dot{\rho}$ and $\dot{\omega}$ (see \eqref{eq:gammadef}), their prediction can be compared to the slope of $\rho$ and $\omega$ immediately after the perturbation.  Fig.~\ref{fig:dynval} shows $\rho_{15}$, $\omega_{15}$, and their predicted initial rate of change (PRoC) based on $\Delta\sigma_{15}$ and $\Delta\gamma_{15}$, respectively.  The exactness of the second-order strength metrics is also verified.

\begin{table}[t]
\caption{Dynamic validation results.}
\begin{center}
\renewcommand{\arraystretch}{1.35}
\begin{tabular}{lccc}
\hline
Variable & Predicted using $\ull{S}'_{15}$ & Read from TDS & Error\\ 
\hline
 $\Delta v_{15}$ (pu)  & -0.00461526 & -0.00461524 & -1.273e-08\\ 
 $\Delta\theta_{15}$ (rad)  &  -0.01730211 & -0.01730211 & 6.545e-09\\
 $\Delta\rho_{15}$ (pu/s) & 0.0 & 2.984e-06 & -2.984e-06\\
 $\Delta\omega_{15}$ (pu/s) & 0.0 & -4.067e-05 & 4.067e-05\\
\hline
\end{tabular}
\end{center}
\label{tab:dynval}
\end{table}

\subsection{Strength of GFL converters}

\begin{figure}[hbtp]
    \centering
    \begin{subfigure}[]{0.49\linewidth}
        \caption{$v_{15}$ and its PRoC.}
        \label{fig:v_15mod}
        \includegraphics[width=0.92\linewidth]{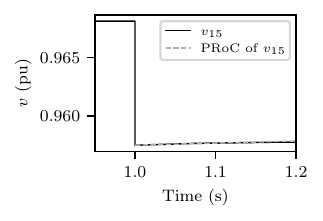}
    \end{subfigure}
        \begin{subfigure}[]{0.49\linewidth}
        \caption{$\theta_{15}$ and its PRoC.}
        \includegraphics[width=0.92\linewidth]{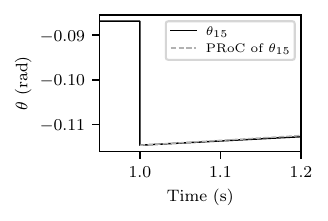}
    \end{subfigure}
    \hfill
    \begin{subfigure}[]{0.49\linewidth}
        \caption{$\rho_{15}$ and its PRoC.}
        \includegraphics[width=0.95\linewidth]{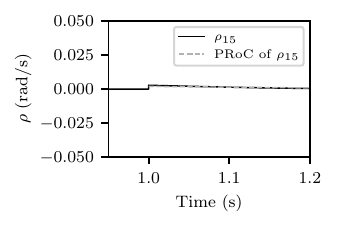}
        \label{fig:rho_15mod}
    \end{subfigure}
        \begin{subfigure}[]{0.49\linewidth}
        \caption{$\omega_{15}$ and its PRoC.}
        \label{fig:om_15mod}
        \includegraphics[width=0.95\linewidth]{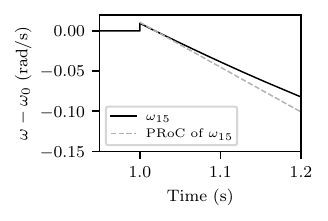}
    \end{subfigure}
    \caption{Trajectories of $\bar{v}_{15}$, $\bar{\eta}'_{15}$ and their PRoC in the modified system.}
    \label{fig:dynvalmod}
\end{figure}

The system is modified by replacing the SMs marked with dashed rectangles in Fig.~\ref{fig:slds} by GFL converters using the model presented in Section \ref{subsec:gfl}.  For this modified system and the same perturbation simulated before, Fig.~\ref{fig:dynvalmod} shows the trajectories of $v_{15}$, $\theta_{15}$, $\rho_{15}$, and $\omega_{15}$ and their respective PRoC calculated using our strength formulation.  A relevant difference with respect to the original system is that $\ull{\bfg A}'$ is no longer null due to the $\ull{a}'$ component of GFLs (see \eqref{eq:gfl_firstorder}).  This implies that $\bar{\eta}'$ is not continuous anymore, i.e., the CF at buses can now experience `jumps', as demonstrated in Fig.~\ref{fig:rho_15mod} and Fig.~\ref{fig:om_15mod}.  The results again verify the accuracy in the prediction of $\Delta\bar{v}_{15}$, $\Delta\bar{\eta}'_{15}$ and $\Delta\bar{\eta}''_{15}$.

\subsection{Strength of GFM converters}

\begin{figure}[hbtp]
    \centering
    \begin{subfigure}[]{0.49\linewidth}
        \caption{$v_{15}$ and its PRoC.}
        \label{fig:v_15mod2}
        \includegraphics[width=0.92\linewidth]{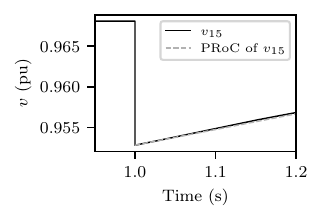}
    \end{subfigure}
        \begin{subfigure}[]{0.49\linewidth}
        \caption{$\theta_{15}$ and its PRoC.}
        \includegraphics[width=0.92\linewidth]{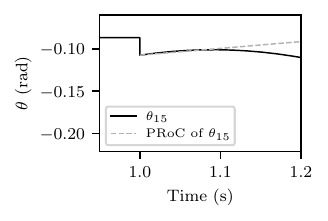}
    \end{subfigure}
    \hfill
    \begin{subfigure}[]{0.49\linewidth}
        \caption{$\rho_{15}$ and its PRoC.}
        \includegraphics[width=0.95\linewidth]{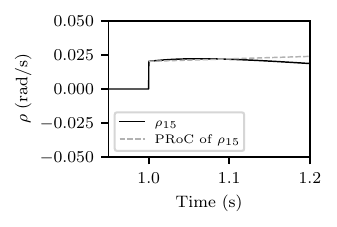}
        \label{fig:rho_15mod2}
    \end{subfigure}
        \begin{subfigure}[]{0.49\linewidth}
        \caption{$\omega_{15}$ and its PRoC.}
        \label{fig:om_15mod2}
        \includegraphics[width=0.95\linewidth]{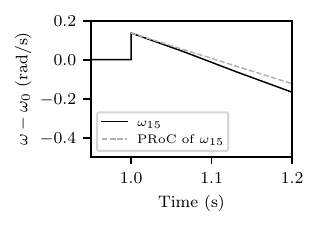}
    \end{subfigure}
    \caption{Trajectories of $\bar{v}_{15}$, $\bar{\eta}'_{15}$ and their PRoC in the modified system.}
    \label{fig:dynvalmod2}
\end{figure}

The system is modified one last time by replacing the three remaining SMs by GFM converters using the model presented in Section \ref{subsec:gfm}. The resulting network is composed entirely by IBRs, seven GFL units and three GFM units. Fig. \ref{fig:dynvalmod2} compares the trajectories of $v_{15}$, $\theta_{15}$, $\rho_{15}$, and $\omega_{15}$ with their respective PRoC calculated using our strength formulation. Once again, the results verify the accuracy in the prediction of $\Delta\bar{v}_{15}$, $\Delta\bar{\eta}'_{15}$ and $\Delta\bar{\eta}''_{15}$. Instead of being dominated by the physics of SMs, the dynamic response of the frequency is driven by the IBRs' control. In this scenario, $\omega$ experiences significantly higher `jumps' and higher rates of change than in presence of SMs, as demonstrated in Fig. \ref{fig:om_15mod2} when compared against Fig. \ref{fig:om_15}  and Fig. \ref{fig:om_15mod}. This behavior is a consequence of the power-frequency droop synchronization mechanism of the GFMs, reflected in the first- and second-order components of them (see (\ref{eq:gfm_zeroorder})-(\ref{eq:gfm_secondorder})). Consistently, the strength metric values are higher in this case, reflecting a weaker, i.e., more sensible, CF---and its rate of change---to sudden changes in the current. It is important to remark that this does not imply the system response has worsen in terms of stability, since strength and stability are different concepts fundamentally decoupled. Our framework demonstrates its ability to accurately quantify strength across all scenarios.

\subsection{Remarks}

% IgP: 
% Paragraph remarking the theoretical scope of the paper.

%However, further research is required to implement our framework in the real world. 

% Below is a list of some of the practical aspects left for future work that are currently work in progress:

\subsubsection{Applications of the proposed framework}

The theoretical foundations of the definition of ``strength'' provided in this paper have a natural next step: the evaluation of typical ranges and sensitivities for each of the proposed twelve metrics.  In this regard, we believe the best approach is statistical and should be based on a real-world power system, realistic operating conditions, and credible contingencies.  This will allow identifying security thresholds as well as the sensitivity and information carried by each metric.  We expect the results to be system-dependent.

% and goes thus beyond hte scope of the current work.

% Assess computational complexity of implemeting the framework in real case scenarios. \textbf{IgP: Clarity off-line nature of calculation done by TSOs here? (to answer R4.Q6).}

% Find correlations between our strength metrics and specific stability issues.  Compare against existing well-accepted strength metrics.  Establish and justify a set of simplifications to make the framework more practical.  

\subsubsection{On the link between stability and the proposed definition of strength}

The proposed framework to evaluate system strength must not be confused with a surrogate to system stability analysis.  System strength and system stability are two different concepts that, although related in some cases (e.g., a weak system can be more easily prone to instability), are fundamentally decoupled.  The study of the link between the metrics proposed in this paper and the occurrence of instabilities in the system requires a thorough, dedicated study, which will be collected in a follow-up work.
  
\subsubsection{On the computational burden of the proposed framework}

Given the offline nature of system strength analysis, the computational burden of the proposed metrics is neither essential to them nor to power system strength in general. In any case, all calculations can exploit the fact that all factorization operations involved are on sparse matrices. Calculating the full set of metrics for each scenario in this case study takes a few seconds, and for a dynamic model of one order of magnitude larger, the analysis requires about 30 seconds per contingency.  Moreover, because contingencies are studied independently from each other, the calculation of the metrics can be easily parallelised.

\section{Conclusions}
\label{sec:conclusion}

% \subsection{Conclusions}
This paper establishes the theoretical foundations for a general and unifying framework for power system strength.  
% The concept of strength is conceived as a property of the voltage at buses that represents its resistance to perturbations, where `voltage' refers to the multidimensional electrical variable that is in motion in AC systems.  In practice, 
The formulation features a set of twelve indicators organized into three different dynamical orders that capture the voltage strength when subjected to changes in current injections.

The paper also presents a systematic approach to studying the impact of different devices on strength and provides examples featuring basic models. For instance, the key parameters of SMs are the internal reactance, which mostly affects the zero-order strength, and the inertia, which dominates the second-order strength. Furthermore, a network composed exclusively of SMs forces the first-order CF to be continuous, a characteristic lost in the presence of GFLs due to their first-order component. In a 100\% IBR system, composed of GFLs and power-frequency droop GFMs, the CF and its rate of change experience higher sudden changes, dictated by the first- and second-order components of their controllers. Our formulation accurately quantifies strength across all scenarios. Therefore, it is not limited to any specific combination of devices composing the system, nor to any imperative dynamics.

% \color{purple}

% \subsection{Future work}
% IgP: 
% Paragraph remarking the theoretical scope of the paper. However, further research is required to implement our framework in the real world. 

Future work will focus on practical aspects and applications of the proposed theoretical framework, incuding the comparison of the proposed framwork with existing metrics, and the study of the link between the proposed metrics and the stability of the system.

\vspace{-6mm}

\begin{IEEEbiography}[{\includegraphics[width=1in, height=1.25in, clip, keepaspectratio]{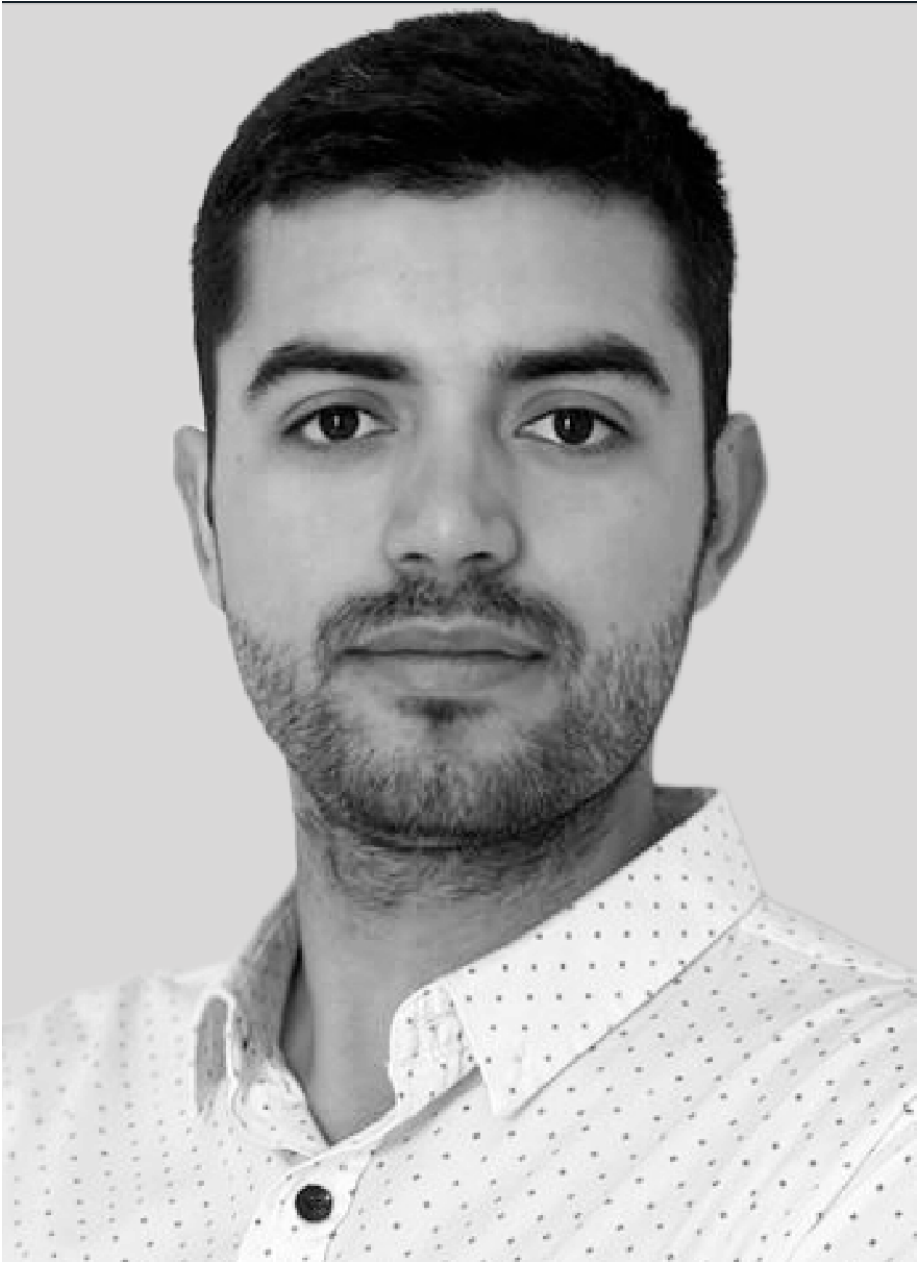}}] {Ignacio Ponce} received from the University of Chile the BSc.~and MSc.~degree in Electrical Engineering in 2019 and 2022, respectively.  He is currently pursuing a Ph.D in Electrical Engineering at University College Dublin, Ireland.  His research interests include power system modeling, control and stability analysis.
\end{IEEEbiography}

\vspace{-6mm}

\begin{IEEEbiography}[{\includegraphics[width=1in, height=1.25in, clip, keepaspectratio]{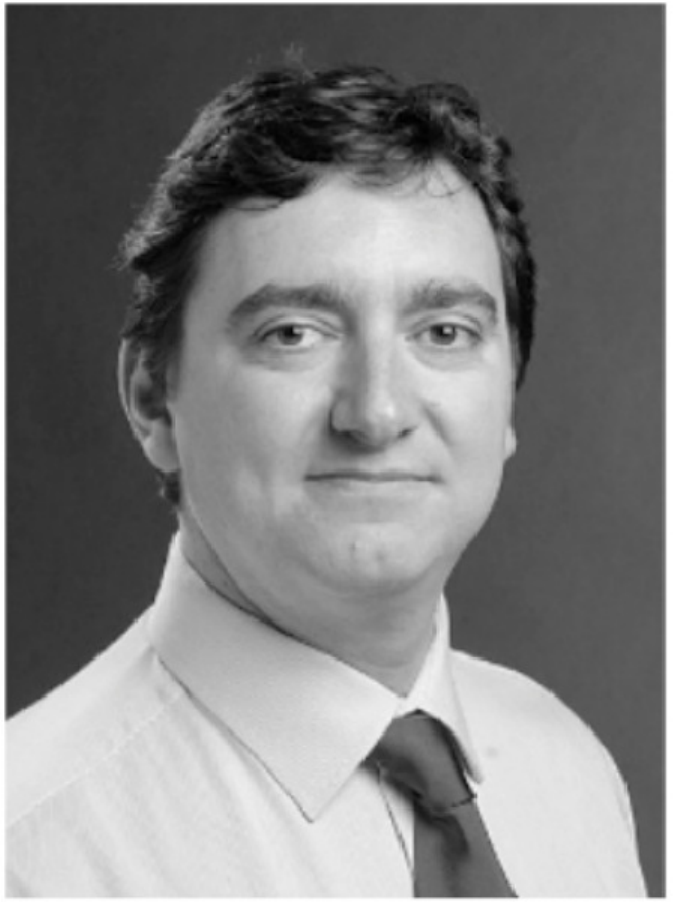}}] {Federico Milano} (F'16) received from the Univ.~of Genoa, Italy, the Ph.D.~in Electrical Engineering 2003.  In 2013, he joined the University College Dublin, Ireland, where he is currently a full professor.  He is Chair of the IEEE Power System Stability Controls Subcommittee, IET Fellow, IEEE PES Distinguished Lecturer, Senior Editor of the IEEE Transactions on Power Systems, Member of the Cigr{\'e} Irish National Committee, and Co-Editor in Chief of the IET Generation, Transmission \& Distribution.  His research interests include power system modeling, control and stability analysis.
\end{IEEEbiography}

\vfill

\end{document}